\documentclass[twocolumn,twocolappendix]{aastex63}
\usepackage{ragged2e}
\usepackage{natbib}

\accepted{ApJ}

\shorttitle{Ionized Outflows in Quasar Host Galaxies}
\shortauthors{Molina et al.}

\begin{document}

\title{Ionized Outflows in Nearby Quasars are Poorly Coupled to their Host Galaxies}

\correspondingauthor{Ran Wang}
\email{rwangkiaa@pku.edu.cn}

\author[0000-0002-8136-8127]{Juan Molina}
\affil{Kavli Institute for Astronomy and Astrophysics, Peking University, Beijing 100871, China}

\author[0000-0001-6947-5846]{Luis C. Ho}
\affil{Kavli Institute for Astronomy and Astrophysics, Peking University, Beijing 100871, China}
\affiliation{Department of Astronomy, School of Physics, Peking University, Beijing 100871, China}

\author[0000-0003-4956-5742]{Ran Wang}
\affil{Kavli Institute for Astronomy and Astrophysics, Peking University, Beijing 100871, China}
\affiliation{Department of Astronomy, School of Physics, Peking University, Beijing 100871, China}

\author[0000-0002-4569-9009]{Jinyi Shangguan}
\affil{Max-Planck-Institut f\"{u}r Extraterrestrische Physik (MPE), Giessenbachstr., D-85748 Garching, Germany}

\author[0000-0002-8686-8737]{Franz E. Bauer}
\affil{Instituto de Astrof{\'{\i}}sica and Centro de Astroingenier{\'{\i}}a, Facultad de F{\'{i}}sica, Pontificia Universidad Cat{\'{o}}lica de Chile, Casilla 306, Santiago 22, Chile}
\affiliation{Millennium Institute of Astrophysics (MAS), Nuncio Monse{\~{n}}or S{\'{o}}tero Sanz 100, Providencia, Santiago, Chile} \affiliation{Space Science Institute, 4750 Walnut Street, Suite 205, Boulder, Colorado 80301}

\author[0000-0001-7568-6412]{Ezequiel Treister}
\affil{Instituto de Astrof{\'{\i}}sica and Centro de Astroingenier{\'{\i}}a, Facultad de F{\'{i}}sica, Pontificia Universidad Cat{\'{o}}lica de Chile, Casilla 306, Santiago 22, Chile}

\author[0000-0001-5105-2837]{Ming-Yang Zhuang}
\affil{Kavli Institute for Astronomy and Astrophysics, Peking University, Beijing 100871, China}
\affiliation{Department of Astronomy, School of Physics, Peking University, Beijing 100871, China}

\author{Claudio Ricci}
\affil{N{\'{u}}cleo de Astronom{\'{\i}}a de la Facultad de Ingenier{\'{\i}}a Universidad Diego Portales, Av. Ej{\'{e}}rcito Libertador 441, Santiago 22, Chile}
\affiliation{Kavli Institute for Astronomy and Astrophysics, Peking University, Beijing 100871, China}

\author{Fuyan Bian}
\affil{European Southern Observatory, Alonso de C{\'{o}}rdova 3107, Casilla 19001, Vitacura, Santiago 19, Chile}
\affiliation{Research School of Astronomy \& Astrophysics, Mt Stromlo Observatory, Australian National University, Canberra, ACT 2611, Australia}

\begin{abstract}
We analyze Multi Unit Spectroscopic Explorer observations of nine low-redshift ($z < 0.1$) Palomar-Green quasar host galaxies to investigate the spatial distribution and kinematics of the warm, ionized interstellar medium, with the goal of searching for and constraining the efficiency of active galactic nucleus (AGN) feedback.  After separating the bright AGN from the starlight and nebular emission, we use pixel-wise, kpc-scale diagnostics to determine the underlying excitation mechanism of the line emission, and we measure the kinematics of the narrow-line region (NLR) to estimate the physical properties of the ionized outflows. The radial size of the NLR correlates with the AGN luminosity, reaching scales of $\sim 5$ kpc and beyond.  The geometry of the NLR is well-represented by a projected biconical structure, suggesting that the AGN radiation preferably escapes through the ionization cone. We find enhanced velocity dispersions ($\gtrsim 100$\,km\,s$^{-1}$) traced by the H$\alpha$ emission line in localized zones within the ionization cones. Interpreting these kinematic features as signatures of interaction between an AGN-driven ionized gas outflow and the host galaxy interstellar medium, we derive mass outflow rates of $\sim 0.008-1.6\,M_\odot$\,yr$^{-1}$ and kinetic injection rates of $\sim 10^{39}-10^{42}$\,erg\,s$^{-1}$, which yield extremely low coupling efficiencies of $\lesssim 10^{-3}$. These findings add to the growing body of recent observational evidence that AGN feedback is highly ineffective in the host galaxies of nearby AGNs.
\end{abstract}


\section{Introduction}

The accretion of matter onto supermassive black holes (BHs) residing within the nuclei of galaxies is widely accepted to be the source of the prodigious energy released in active galactic nuclei (AGNs; \citealt{Rees1984}). The net amount of energy emitted during the growth of the BH greatly exceeds the binding energy of its host galaxy \citep{Fabian2012}, positioning the AGN-host galaxy interplay as one of the key elements that influences the evolution of galaxies (e.g., \citealt{Ho2004,Heckman2014,Harrison2018}). The correlations between the BH mass and the bulge properties of the host \citep{Kormendy1995,Richstone1998,Ferrarese2000,Gebhardt2000,Tremaine2002,Kormendy2013} suggest that the possible coupling between the growth of the BH and its host galaxy occurs on nuclear scales (e.g., \citealt{HopkinsElvis2010}), where AGN-driven winds may interact with gas to produce multi-phase outflows (e.g., \citealt{Cicone2014,Feruglio2015,Karouzos2016,Fiore2017,Morganti2017,Cicone2018,Fluetsch2019}) on galaxy-wide scales that are powerful enough to heat or expel the interstellar medium (ISM) from the host galaxy (e.g., \citealt{Silk1998,Somerville2008,Schaye2015,Sijacki2015,Lacey2016}), thereby suppressing star formation \citep{Dubois2016} and keeping the host galaxy quiescent \citep{Fabian2012}.

In the context of the geometric structure envisioned by the unified model of AGNs \citep{Antonucci1993}, radiation pressure would clear portions of the dusty toroidal region surrounding the accretion disk and broad-line region (BLR; \citealt{Ricci2017}), allowing the escape of ionizing photons. The AGN radiation permeates through the host galaxy disk, likely in the form of biconical ionization zones, forming the so-called narrow-line region (NLR; e.g., \citealt{Haniff1988,Pogge1988,Wilson1994}). The NLR extension increases with AGN luminosity \citep{Bennert2002,Schmitt2003a,Schmitt2003b,Netzer2004,Greene2012,Liu2013,Liu2014,Hainline2014,Storchi2018,Chen2019,Husemann2019b}, and the AGN ionizing photons can even reach regions beyond the host galaxy if there is gas to be ionized (e.g., \citealt{Wampler1975,Stockton1976,Stockton1987,Lintott2009,Fu2009,Keel2012,Kreimeyer2013,Schweizer2013,Watkins2018}).\footnote{While some recent studies use the term ``extended narrow-line region'' (ENLR),  we prefer to use the more traditional terminology of ``NLR'' and reserve ``ENLR'' to refer to gaseous regions outside of the host galaxy that are ionized by the AGN radiation (e.g., \citealt{Fu2009}).} However, the outflow signatures in ionized gas tend to be observed in the innermost zones of the NLR (e.g., \citealt{Schnorr2014,Riffel2015,SunAGN2017,Fischer2018,Kang2018,Husemann2019b}).

Despite the ubiquitous evidence of multi-phase gas outflows in AGN hosts, the role of AGN feedback in establishing the host galaxy properties and its ability to quench ongoing star formation activity are still intensively debated. Local low-luminosity AGNs exhibit global star formation rates (SFRs) lower than those measured from the inactive, star-forming galaxy population \citep{Ellison2016,Leslie2016,Jackson2020}, and some systems are effectively passive \citep{Ho2003}. On the other hand, the more luminous nearby AGN host galaxies are similar to the inactive galaxy population in several key properties, such as SFR, atomic hydrogen gas content, molecular gas mass, and dust emission and/or dust attenuation \citep{Maiolino1997,Evans2001,Scoville2003,Evans2006,Bertram2007,Ho2008,Fabello2011,Harrison2012,Gereb2015,Zhu2015,Husemann2017,Bernhard2019,Ellison2019AGN,Shangguan2018,Shangguan2019,Shangguan2020,Grimmett2020,Jarvis2020,Yesuf2020,Zhuang2021}. Far from being quenched, stars are actively forming within the host galaxies of luminous AGNs \citep{Shangguan2018}, and in some the star formation efficiencies are comparable to those seen in starburst systems \citep{Kirkpatrick2020,Shangguan2020b,Zhuang2020,Xie2021}. These observations suggest that negative AGN feedback, if present, likely produces only a localized effect on the host galaxies.

\begin{figure}
\centering
\includegraphics[width=0.9\columnwidth]{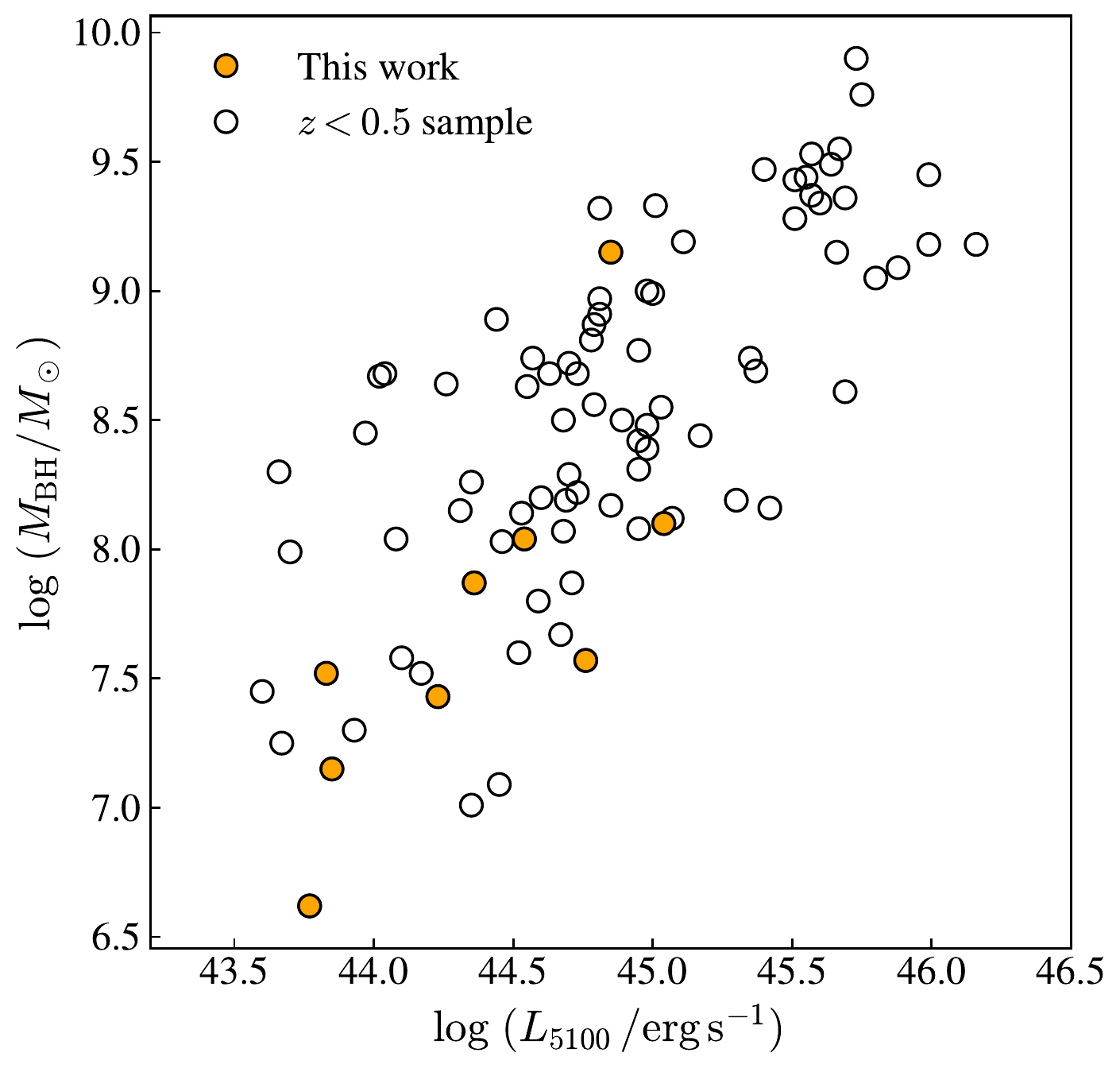}
\caption{\label{fig:PG_sample} Black hole mass as a function of the AGN monochromatic luminosity at 5100\,\r{A} for the PG quasars.  The open circles show the sample at $z < 0.5$ \citep{Shangguan2018}. We focus on the nearest ($z < 0.1$) host galaxies with available MUSE data. These targets mostly sample the $L_{5100}  \lesssim 10^{45}$\,erg\,s$^{-1}$ and $M_{\rm BH} \lesssim 10^8$\,$M_\odot$ range, with PG\,1426+015 being the only exception (Table~\ref{tab:sample}).}
\end{figure}

\begin{table*}
	\centering
	\def\arraystretch{1.2}
	\setlength\tabcolsep{3pt}
    	\caption{\label{tab:sample} Basic Parameters of the Sample}
    	\vspace{0.2mm}
	\begin{tabular}{ccccccccccc}
		\hline
		\hline
		Object & R.A. & Decl. & $z$ & $D_L$ & Morphology & $\log M_\star$ & $\log L_{\rm IR}$ & SFR$_{\rm IR}$ & $\log M_{\rm BH}$ & $\log L_{5100}$ \\
		& (J2000.0) & (J2000.0) & & (Mpc) & &($M_\odot$) & (erg\,s$^{-1}$) & ($M_\odot$\,yr$^{-1}$) & ($M_\odot$) & (erg\,s$^{-1}$)\\
		(1) & (2) & (3) & (4) &(5) & (6) & (7) & (8)& (9) & (10) & (11) \\
		\hline
	PG\,0050+124 & 00:53:34.94 & +12:41:36.2 & 0.061 & 282.3 & Disk & 11.12 & 44.94$^{+0.01}_{-0.01}$ & 26.3$^{+0.7}_{-0.5}$ & 7.57 & 44.76\\
         PG\,0923+129 & 09:26:03.29 & +12:44:03.6 & 0.029 & 131.2 & Disk & 10.71 & 44.05$^{+0.01}_{-0.02}$ & 3.4$^{+0.1}_{-0.2}$ & 7.52 & 43.83\\
         PG\,0934+013	& 09:37:01.05 & +01:05:43.2 & 0.051 & 229.6 & Disk & 10.38 & 43.96$^{+0.02}_{-0.02}$ & 2.7$^{+0.1}_{-0.1}$ & 7.15 & 43.85\\
         PG\,1011$-$040 & 10:14:20.69 & $-$04:18:40.5 & 0.058 & 267.9 & Disk & 10.87 & 43.98$^{+0.02}_{-0.02}$ & 2.9$^{+0.2}_{-0.2}$ & 7.43 & 44.23\\
         PG\,1126$-$041 & 11:29:16.66 & $-$04:24:07.6 & 0.060 & 277.5 & Disk & 10.85 & 44.46$^{+0.03}_{-0.03}$ & 8.7$^{+0.7}_{-0.6}$ & 7.87 & 44.36\\
         PG\,1211+143 & 12:11:17.67 & +14:03:13.2 & 0.082 & 400.0 & Disk & 10.38 & 43.32$^{+0.05}_{-0.05}$ & 0.6$^{+0.08}_{-0.07}$ & 8.10 & 45.04\\
        PG\,1244+026 & 12:46:35.25  & +02:22:08.8 & 0.048 & 220.1 & Disk & 10.19  & 43.85$^{+0.02}_{-0.01}$ & 2.1$^{+0.1}_{-0.1}$ & 6.62 & 43.77\\
         PG\,1426+015 & 14:29:06.57 & +01:17:06.2 & 0.087 & 400.5 & Merger & 11.05 & 44.55$^{+0.02}_{-0.02}$ & 10.8$^{+0.5}_{-0.5}$ & 9.15 & 44.85\\
         PG\,2130+099 & 21:32:27.81 & +10:08:19.5 & 0.061 & 292.3 & Disk & 10.85 & 44.37$^{+0.02}_{-0.03}$ & 7.1$^{+0.4}_{-0.5}$ & 8.04 & 44.54\\
		\hline
	\end{tabular}
	\justify
	{\justify \textsc{Note}--- (1) Source name. (2) Right ascension. (3) Declination. (4) Redshift. (5) Luminosity distance. (6) Morphological type of the host galaxy \citep{Zhang2016,Kim2017,Zhao2021} (7) Stellar mass, with $1\,\sigma$ uncertainty of 0.3\,dex \citep{Shangguan2018}. (8) Total infrared luminosity of the host galaxy \citep{Shangguan2018}. (9) Infrared-based SFR estimated by adopting Equation~4 of \citet{Kennicutt1998b} and a \cite{Kroupa2001} stellar initial mass function. (10) Black hole mass, estimated by applying the calibration of \citet{Ho2015}, taken from \citep{Shangguan2018}. (11) AGN monochromatic luminosity at 5100\,\r{A}. }
\end{table*}

The most luminous AGNs, the quasars, are the ideal laboratories to scrutinize the interplay between AGN feedback and the host galaxy ISM. Within the context of the merger-induced evolution scenario of quasars \citep{Sanders1988}, the birth of a largely unobscured quasar is produced after the overwhelming release of energy by the AGN expels the enshrouding gas and dust near the nucleus \citep{Hopkins2008,Treister2010}. However, studying active galaxies is challenging because the radiation from the AGN contaminates the emission from the underlying host, a problem that is most severe for type~1 (unobscured, broad-line) sources, although it is not entirely negligible even in type~2 (obscured, narrow-line) systems. Integral-field unit (IFU) observations offer the possibility to overcome such a limitation by spatially resolving the host galaxies (e.g., \citealt{Jahnke2004,Lipari2009,Husemann2014,Harrison2016,Husemann2017,Ilha2019,Feruglio2020,Kakkad2020,Lacerda2020,Riffel2021,Scholtz2021}). The AGN emission, as a point-like source that follows the observation point-spread function (PSF), in principle can be deblended from the extended emission of the host using sophisticated computational procedures (e.g., \citealt{Husemann2013,Rupke2017,Husemann2022}) to produce cleaner measurements of the host galaxy properties.

This work reports on seeing-limited and ground-layer adaptive optics-aided (FWHM seeing $\sim 0\farcs6-1\arcsec$, which corresponds to physical scales $\lesssim 1$ kpc) Very Large telescope (VLT) Multi Unit Spectroscopic Explorer (MUSE) observations targeting the optical light toward the host galaxies of nine Palomar-Green (PG) quasars \citep{Schmidt1983}. The low-redshift ($z < 0.5$) subset of the PG survey contains 87 type~1 quasars \citep{Boroson1992}. This sample is one of the most studied quasar surveys to date, enjoying a rich repository of multi-wavelength data for the AGN and host galaxy, encompassing X-ray observations \citep{Reeves2000,Piconcelli2005,Bianchi2009}, optical long-slit spectra \citep{Boroson1992,Ho2009}, Hubble Space Telescope (HST) high-resolution imaging of the host galaxy \citep{Kim2008,Kim2017,Zhang2016,Zhao2021}, dust properties for both the torus and host galaxy \citep{Shi2014,Petric2015,Shangguan2018,Zhuang2018}, global SFRs (\citealt{Xie2021}), molecular gas masses measured from CO \citep{Shangguan2020,Shangguan2020b}, and radio properties \citep{Kellermann1989,Kellermann1994}.

The main goal of this study is to characterize the effects of AGN feedback on the host galaxy traced by optical nebular emission lines.  We focus on a small subset of the nearest ($z < 0.1$) PG quasars with $L_{5100} \equiv \lambda L_\lambda(5100\,\rm \AA) \lesssim 10^{45}$\,erg\,s$^{-1}$ and $M_{\rm BH} \lesssim 10^8$\,$M_\odot$ (Figure~\ref{fig:PG_sample}). The upper limit on redshift is chosen to avoid mixing observations with large differences in physical scale resolution, but trading off the quasar luminosity and BH mass range covered by the sample. We employ no further constraints to select our targets.  A companion publication will examine the star formation activity and star formation efficiency of the host galaxies. 

\section{Sample and Observations}
\label{sec:obs}

We analyze optical light VLT-MUSE wide-field-mode IFU observations of nine low-redshift ($z < 0.1$) PG quasar host galaxies (Table~\ref{tab:sample}). We observed three targets during our European Southern Observatory (ESO) program 0103.B$-$0496(B; PI:~F.~Bauer). We complement those observations with archival VLT-MUSE data of PG quasar host galaxies taken from the ESO programs 094.B$-$0345(A), 095.B$-$0015(A), 097.B$-$0080(A), 0101.B$-$0368(B), and 0104.B$-$0151(A). All observations were carried out from January 2015 to October 2019. The field-of-view of a single MUSE observation is nearly $1\arcmin\times 1\arcmin$ with a pixel size sampling of $0\farcs2 \times 0\farcs2$, which generates $\sim 90,000$ spectra per pointing. The MUSE spectral coverage ranges from $\sim$4700 to $\sim$9350\,\r{A}, with a wavelength sampling of 1.25\,\r{A}\,channel$^{-1}$ and at a mean full width at half maximum (FWHM) resolution of $R \approx 3000$ (2.65\,\r{A}). We analyze the MUSE archival data cubes,\footnote{http://archive.eso.org/scienceportal/home} which correspond to a combination of multiple observing blocks (OBs) typically ranging from 2 to 4 for single program observations. In the particular case of PG\,0050+124, we used the ``MUSE-DEEP''\footnote{http://www.eso.org/rm/api/v1/public/releaseDescriptions/102} reduced data cube, which corresponds to a combination of 15 single OBs focused on maximizing the output signal contrast. The benchmark is to obtain a better signal-to-noise (S/N) ratio than any individual observing run data cube. MUSE and MUSE-DEEP data cubes are reduced following the same pipeline framework outlined in \citet{Weilbacher2020}. We note that all the observations were carried out in good weather conditions, with most of the OBs graded ``A.''  A few OBs were graded ``B'' because the sky conditions did not meet the requested observation seeing (e.g., PG\,1126$-$041). Only one OB was graded ``C'' because the data reduction pipeline did not converge when performing the master sky line fit\footnote{https://www.eso.org/rm/api/v1/public/releaseDescriptions/78}. This means that only the sky lines were not properly modeled and subtracted due to an inaccurate characterization during the run of the data reduction pipeline. This single OB (from a total of four) pertains to observations of PG\,1426+015, and we included it when analyzing this host galaxy data cube. The observations were taken under average seeing conditions of $\sim 1\arcsec$ (Table~\ref{tab:obssetup}). Five targets were observed under natural seeing conditions, while the remaining four MUSE observations were assisted by the ground-layer adaptive optics (GLAO) module. 

We processed the combined data cubes by applying the Zurich Atmosphere Purge (ZAP) sky subtraction tool (version `2.1.dev'; \citealt{Soto2016}) with \textsc{cfwidthSP}\,$= 5$ (the other parameters were not modified) to remove residual sky subtraction features. We masked all the pixels that were affected by spectra flux saturation along the full spectral range. We also masked the spectra at wavelengths where strong sky-line residuals are present in the original data cubes ($\sim$5578.5, 5894.6, 6301.7, 6362.5, and 7640\,\r{A}) due to the uncertain ZAP correction. We note that in the AO-aided observations the MUSE spectra have been previously masked by the data reduction pipeline at the wavelength range surrounding NaD emission (5840--5940\,\r{A}). The data cube for each source was corrected for Galactic reddening assuming a \citet{Cardelli1989} extinction law and extinction values from \citet{Green2019}. We further checked the flux calibration of the MUSE observations by comparing the host galaxy MUSE-based Sloan Digital Sky Survey (SDSS; \citealt{Abolfathi2018}) $r$- and $i$-band total magnitudes with available SDSS Petrosian magnitudes, finding good agreement between both quantities, with average magnitude difference of $\sim 0.2$ (MUSE magnitudes are lower) and 0.2\,dex scatter. In this work we adopt the MUSE line-spread function (LSF) parametrization of \citet{Guerou2017} to correct the line widths for instrumental resolution.

\begin{table}
	\centering
	\def\arraystretch{1.2}
	\setlength\tabcolsep{2pt}
    	\caption{\label{tab:obssetup} MUSE Observational Setup}
    	\vspace{0.2mm}
	\begin{tabular}{ccccccccccc}
		\hline
		\hline
		Object & Observation & Exposure  & Image & Instrument\\
		& Date & Time\,(s) & Quality\,($\arcsec$) & Mode\\
		(1) & (2) & (3) & (4) & (5)\\
		\hline
		PG\,0050+124 & 06\,Oct.\,2019 & 9180 & 0.83 & WFM-AO\\
		PG\,0923+129 & 28\,Apr.\,2019 & 2440 & 1.49 & WFM-AO\\
         PG\,0934+013 & 04\,Apr.\,2015 & 1350 & 0.81 & WFM-noAO\\
         PG\,1011$-$040 & 15\,Jun.\,2015 & 600 & 1.08 & WFM-noAO\\
         PG\,1126$-$041 & 10\,Sep.\,2015 & 900 & 1.01 & WFM-noAO\\
        	PG\,1211+143 & 01\,Apr.\,2016 & 2800 & 0.76 & WFM-noAO\\
         PG\,1244+026 & 30\,May\,2019 & 2440 & 1.08 & WFM-AO\\
         PG\,1426+015 & 04\,Apr.\,2016 & 2800 & 0.60 & WFM-noAO\\
         PG\,2130+099 & 25\,Sep.\,2019 & 2440 & 0.70 & WFM-AO\\
		\hline
	\end{tabular}
	\justify
{\justify \textsc{Note}--- (1) Source name. (2) Date of the MUSE observations. (3) Total exposure time.
(4) Image quality of the observations. In the case of the AO-aided observations we report the spatial resolution values measured from the white-light image (data cube header keyword ``SKY\_RES''), while for the ``seeing-limited'' observations we report the seeing value. (5) MUSE instrument mode. All the observations were taken in wide-field-mode (WFM), but a few (WFM-AO) were aided by GLAO. Observations taken in natural-seeing conditions (WFM-noAO) were carried out before the GLAO system became operational for MUSE.}
\end{table}

\begin{figure}
\centering
\includegraphics[width=1.0\columnwidth]{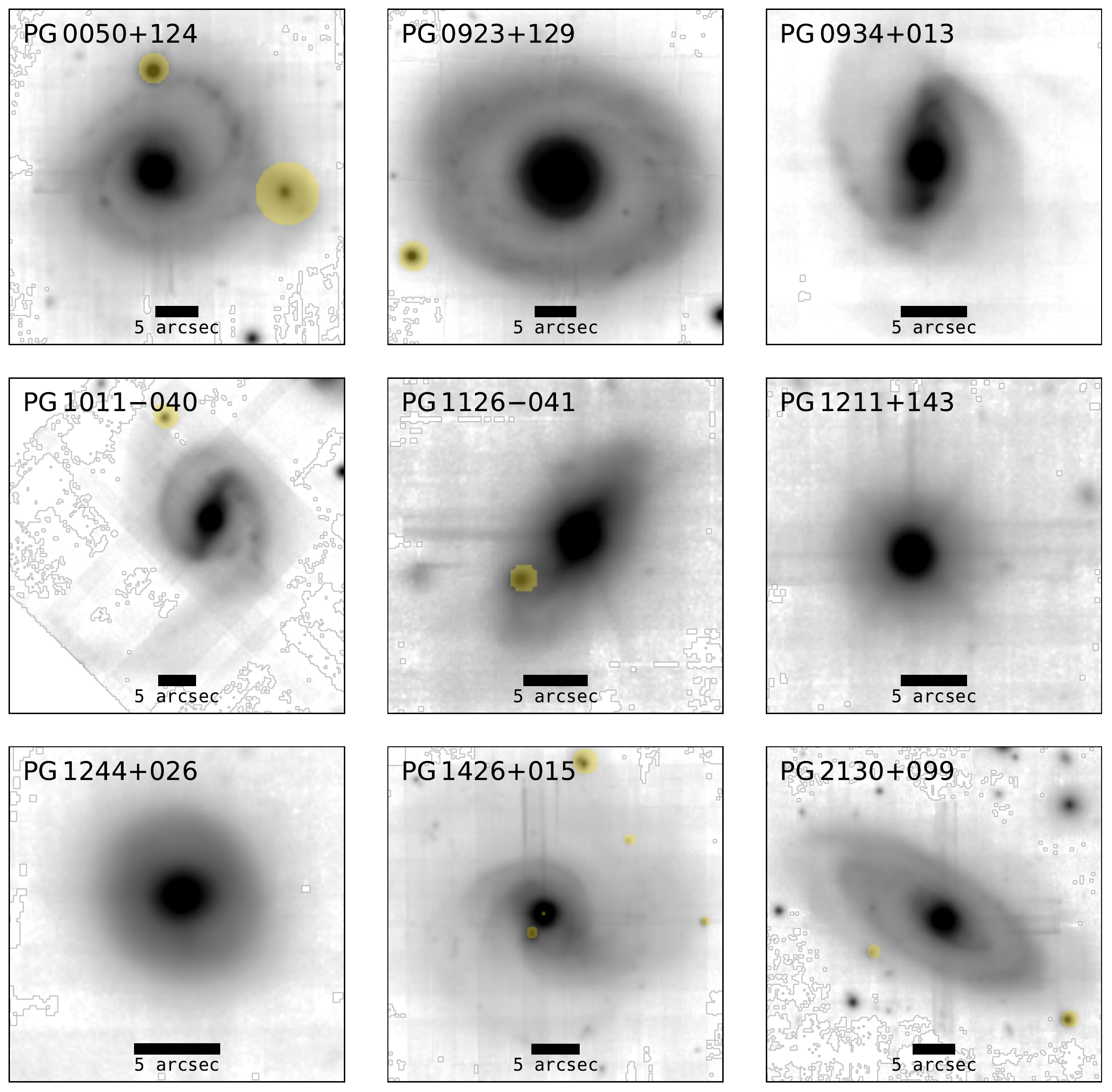}
\caption{\label{fig:White_imgs} MUSE white-light images of the nine PG quasar host galaxies. The yellow-shaded regions represent the circular masks that we manually applied to the individual data cubes to avoid additional sources from contaminating the host galaxy continuum across the line-of-sight.}
\end{figure}

\section{Methods}

Our main goal is to study AGN feedback effects on their host galaxies by analyzing the ionized gas properties. We characterize the host galaxies using two different binning approaches, separately optimized for the stellar component and nebular emission lines. When modeling the host galaxy stellar component, we focus on maximizing the recovery of the S/N of the continuum, even at the expense of degrading the spatial resolution. On the contrary, when characterizing the emission lines we only slightly downgrade the observation spatial resolution to boost the S/N of the data. Preserving the highest possible spatial resolution is critical to deblending the nuclear emission from that of the host, as the nucleus is, by far, the brightest component in the MUSE data cube, and any subtraction error leads to large systematics propagated throughout data analysis. 

As with imaging studies of AGN host galaxies (e.g., \citealt{Zhao2021}), in the IFU data the unresolved emission coming from the nuclear region is seen as a blurred, point-like source that follows the shape of the PSF. This unresolved component needs to be rigorously characterized to not misinterpret the extended emission component. The main difference with respect to the imaging studies is that we have to account for a nuclear spectrum instead of a single flux value per pixel, implying that a correct deblending procedure must include an accurate characterization of this spectrum, which is to be subtracted from the data cube. Two different strategies can be adopted to deblend the AGN emission from that of the host galaxy. The first option is to follow an imaging-based strategy, where the PSF and its variation with wavelength are characterized by modeling pseudo broad-band images derived from the data cube. The model PSF is scaled by the nuclear spectra flux density at each wavelength, and then subtracted from the data cube (e.g., \texttt{QDeblend}$^{\rm 3D}$ \citealt{Husemann2022}). The alternative strategy is to follow a spectra fitting-based approach, where the nuclear spectrum is modeled to then be used as a template when pixel-wise fitting the data cube. This strategy is similar to that of using stellar spectral templates when modeling the galaxy stellar emission. In this work we adopt the latter option as this procedure does not rely on modeling the PSF. Indeed, the PSF is obtained as a by-product because the nuclear spectrum template characterizes the unresolved AGN emission spread over the data cube by the PSF. The nuclear spectrum comprises three components: (1) a featureless continuum from the hot accretion disk, which arises from scales of less than a few light days (e.g., \citealt{Homayouni2019,Guo2022,Jha2022}); (2) broad emission lines from the BLR, which has a size that scales with the luminosity of the source, but even in the most luminous sources studied to date rarely exceeds a light-year (e.g., \citealt{Kaspi2000,Li2021}); and (3) a compact component associated with the photoionized NLR near the vicinity of the nucleus, on scales much smaller than the spatial resolution of our observations ($\sim$\,kpc scale). We emphasize that subtracting the inner NLR component is critical. The [O\,{\sc iii}]\,$\lambda \lambda 4959,5007$ emission produced near the nucleus can achieve a S/N even higher than that for broad H$\beta$, meaning that in every pixel in which we observe the blurred broad H$\beta$ emission we also observe nuclear [O\,{\sc iii}] emission due to the observation finite spatial resolution (see also \citealt{Husemann2016}). Not considering this component would result in an overestimation of the [O\,{\sc iii}] emission and consequently an inaccurate classification of the underlying source powering the nebular emission lines. It would also lead to the confusion of this [O\,{\sc iii}] component as an additional kinematic signature throughout the host galaxy. Keeping in mind this observational effect, we now proceed to explain in detail our data analysis and quasar deblending technique.

\subsection{Stellar Component Modeling}
\label{sec:Stellar-comp}

We first characterize the spatial extent of the host galaxies in the MUSE data cubes by collapsing them across the spectral axis and deriving ``white-light images.'' We use the \textsc{background2D} task of the \textsc{photutils} \textsc{Python} package \citep{Bradley2020} with an $1''$ width smoothing kernel to determine the background level of the white-light images and to mask any emission with S/N\,$<5$. We then use the \textsc{photutils} task \textsc{detect sources} to build segmentation maps and to isolate the host galaxies from other sources within the MUSE field-of-view. At this step, we manually mask any other contaminating object covered by the host galaxy segmentation map sources (Figure~\ref{fig:White_imgs}), including, for instance, small galaxy companions (e.g., PG\,1426+015), field stars (e.g., PG\,2130+099), or both (e.g., PG\,0050+124). 

We further apply Voronoi binning to the host galaxy segmentation map after imposing a target criterion of S/N\,=\,50 \citep{Cappellari2003} at 9010--9280\,\r{A} in the observer frame avoiding sky-line residuals. We employ the penalized pixel-fitting method ({\tt pPXF}; \citealt{Cappellari2004}) and model the spectra of the Voronoi cells using a stellar template library generated from the extended MILES stellar population synthesis models (MIUSCAT; \citealt{Vazdekis2012}). The MIUSCAT composite templates are based on the MILES and calcium triplet (CaT) empirical stellar spectral libraries for single stellar populations (SSPs) of a given age and metallicity, covering the wavelength range 3465--9469\,\r{A} at an uniform spectral resolution of ${\rm FWHM} = 2.51\,$\r{A} (for more details, see \citealt{Vazdekis2012}). The MIUSCAT templates are broadened in velocity space by taking into account their spectral resolution difference with respect to the MUSE LSF width. We mask wide 4800--5100\,\r{A} and 5400--6740\,\r{A} spectral windows in rest-frame and mainly covering the H$\beta$, [O\,{\sc iii}], H$\alpha$, [N\,{\sc ii}]\,$\lambda \lambda 6548,6583$, and [S\,{\sc ii}]\,$\lambda \lambda 6716,6731$ emission lines among others (e.g., He\,{\sc i}\,$\lambda 5875$). We also mask the wavelength ranges 7230--8070\,\r{A} and 8270--8510\,\r{A}, where additional sky lines are observed. We note that in our observations the stellar kinematics are constrained mainly by ${\rm CaT}\,\lambda\lambda 8498, 8542, 8662$ and Mg\,{\sc i}\,$\lambda\lambda 5167, 5173, 5184$.  Owing to the modest S/N and spectral resolution of the MUSE spectra, we only model the absorption lines as Gaussian functions, avoiding more sophisticated fits using a Gauss-Hermite series \citep{Cappellari2004}. 

The {\tt pPXF} procedure allows the inclusion of additive and multiplicative polynomials during the fit, aiming to correct imperfect sky subtraction or scattered light with the former, and inaccuracies in spectral calibration or mismatches in dust reddening correction with the latter \citep{Cappellari2017}. We experimented with both. Our results are insensitive to the inclusion of a multiplicative polynomial of order 3 or less. Conversely, the inclusion of an additive polynomial preferentially downweights the young (age $\lesssim 1\,$Gyr) SSP template, fails to produce sufficiently strong hydrogen absorption lines, and hence introduces systematic biases (overestimates) in the resulting H$\alpha$ and H$\beta$ emission-line fluxes. In light of these considerations, we do not include any additive or multiplicative polynomials during the stellar continuum modeling.

\subsection{Template for the Nuclear Spectra}
\label{sec:Nuc-comp}

We extract the nuclear spectrum for each target by spatially collapsing the data cube within a circular region with radius equal to the observation spatial resolution (Table~\ref{tab:obssetup}), centered on the peak of the quasar emission. Following standard practice (e.g., \citealt{Boroson1992, Ho2009}), we decompose the continuum of the nuclear spectrum into three components: a featureless continuum to represent accretion disk emission, a ``pseudo-continuum'' produced by many blended broad iron emission lines, and starlight from the host galaxy. The featureless continuum is modeled as a power law characterized by an amplitude, spectral index, and pivot wavelength. We constrain the spectral index to be negative, and the pivot wavelength is restricted to lie within 4600--6000\,\r{A} in the rest frame. We use the Fe\,{\sc ii} template of PG\,0050+124 ({\sc I}\,Zw\,I) provided by \cite{Boroson1992} to model the iron pseudo-continuum, allowing it to adjust in amplitude, velocity width, and velocity shift \citep{Hu2008}.

The central stellar component is very poorly constrained because it is outshone by the quasar emission, whose blue, featureless continuum mimics and is degenerate with a young SSP.  To circumvent this difficulty, we adopt an empirical template of the stellar continuum extracted from regions of the host galaxy that are relatively uncontaminated by the AGN. The critical assumption, of course, is that the stellar population at the off-nuclear position is similar to that of the nuclear position. We quantify the AGN contamination as a function of radius by studying the radial variation of the flux density at $\sim 4650\,$\r{A}. Assuming that the MUSE PSF can be described by a \citet{Moffat1969} function with index $\beta = 2.2$ (e.g., \citealt{Bacon2017,Guerou2017}) and using the image quality values recorded during the observations (Table~\ref{tab:obssetup}), we estimate that the AGN continuum on average drops to $\sim 4$\,\% (range $\sim 3\,\%-7$\,\%) of its central peak value\footnote{Note that the level of AGN contamination should be even lower at redder wavelengths because of the decrease of the quasar featureless continuum and the MUSE PSF FWHM variation with wavelength.} at a projected distance of 5 times the seeing radius. Hence, for each host galaxy, we extract an AGN-free spectrum at an annulus centered at 5 times the seeing radius from the nucleus. The best-fit {\tt pPXF} model then serves as the starlight template for the nuclear spectrum.

\begin{figure*}
\centering
\includegraphics[width=1.8\columnwidth]{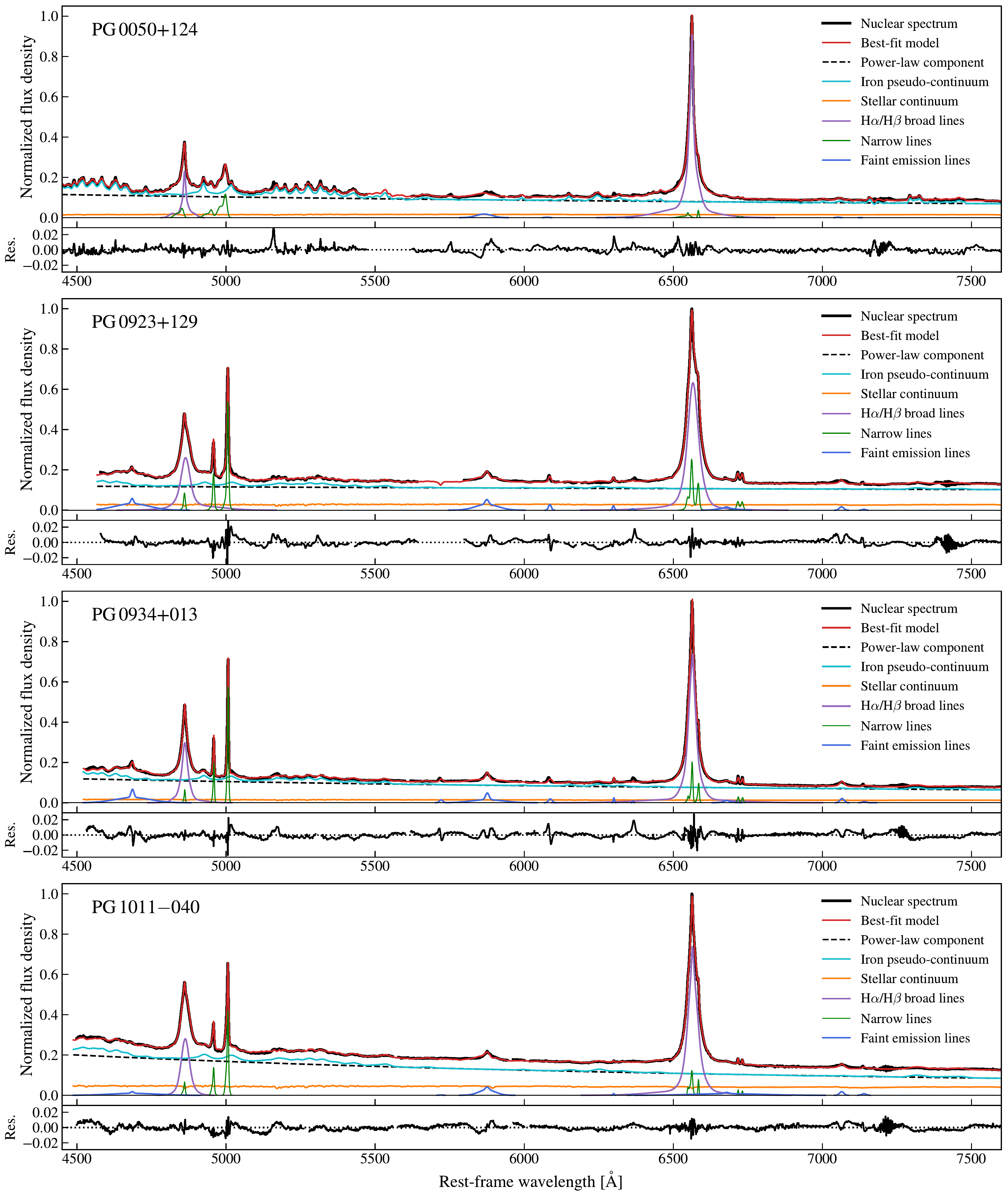}
\caption{\label{fig:qso_spectra_model_a} Nuclear spectra for the nine PG quasar host galaxies. The spectra are extracted from the MUSE data cubes using a circular region centered on the quasar location and with aperture equal to the observation spatial resolution (Table~\ref{tab:obssetup}). We only show the wavelength range covered by our spectral modeling routine. The flux densities are normalized with respect to the peak value of the H$\alpha$ emission. The iron pseudo-continuum component is shifted above the power-law component to improve visualization. The broad-line components correspond to the sum of all the Gaussian sub-components (typically four in each case). The residuals (data$\,-\,$model) are below the $\sim$2.5\,\% limit in all cases.}
\end{figure*}

\begin{figure*}
\centering
\includegraphics[width=1.8\columnwidth]{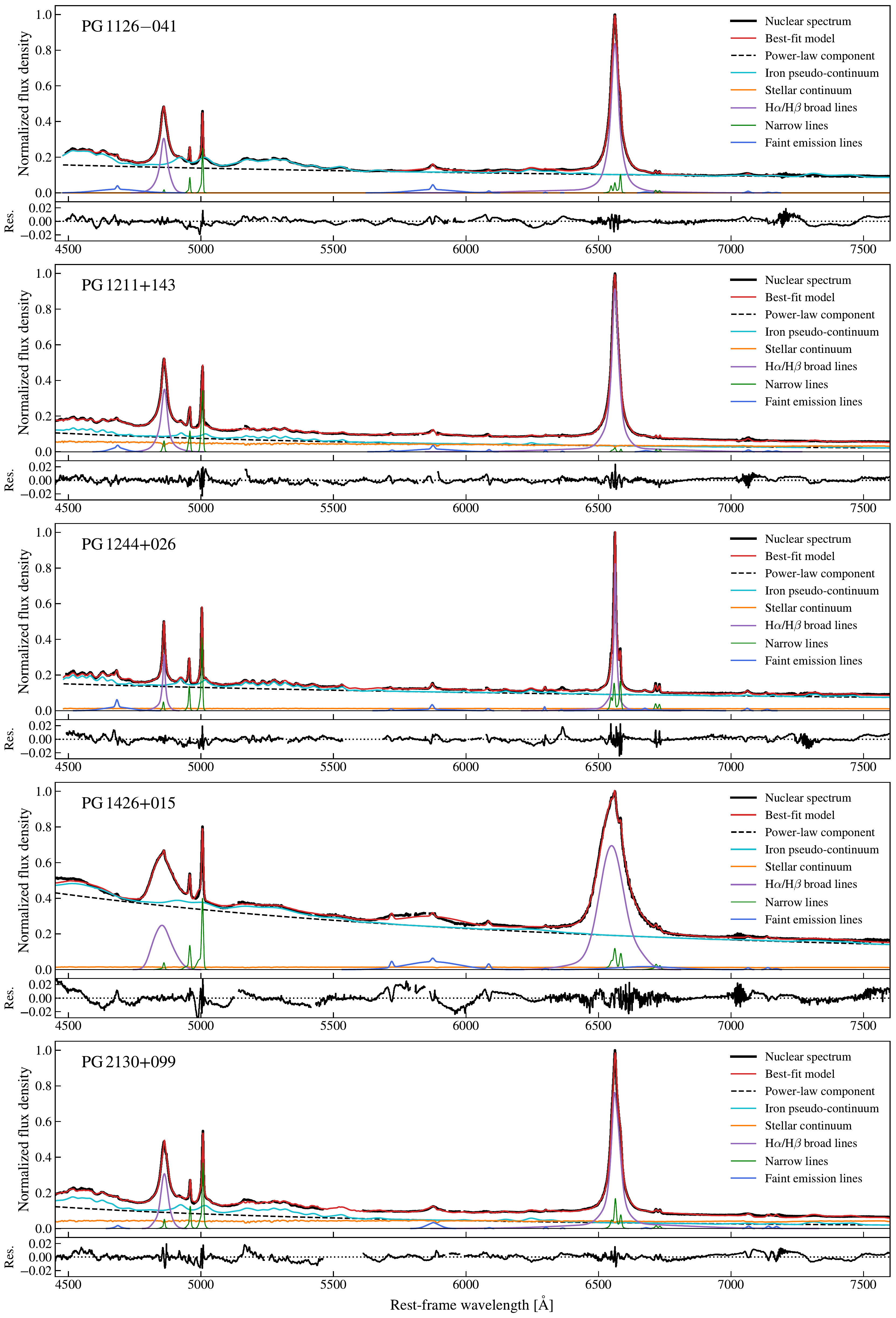}\\
\textbf{Figure~\ref{fig:qso_spectra_model_a} continued.}
\end{figure*}

The emission-line spectrum contains blends of many features, which need to be decomposed carefully. The permitted lines (e.g., H$\beta$ and H$\alpha$) are especially complicated because they comprise heavily blended contributions from the BLR and NLR, each of which can have complex kinematics. Following common practice (e.g., \citealt{Greene2005b}), we first fit the well-isolated, strong narrow forbidden lines of the [O\,{\sc iii}]\,$\lambda \lambda 4959,5007$ doublet, fixing their known separation and relative intensity \citep{Osterbrock2006}, and assuming that both components share the same profile. The lines, though narrow, often have non-trivial profiles.  We model them using as many Gaussians as necessary to achieve a satisfactory fit, although in practice two suffice: a narrow core plus a broader, often blueshifted wing. We then use the best-fitting model for [O\,{\sc iii}]\,$\lambda5007$ as an empirical template to extract the flux of the narrow component of H$\beta$ and H$\alpha$, as well as [N\,{\sc ii}]\,$\lambda \lambda 6548, 6584$, which normally is also severely blended with H$\alpha$. Note that in principle [S\,{\sc ii}]\,$\lambda \lambda 6716, 6731$ gives a more suitable template for treating the H$\alpha$+[N\,{\sc ii}] complex \citep{Ho1997}, but in our sample [S\,{\sc ii}] has much lower S/N than [O\,{\sc iii}], and we use the [O\,{\sc iii}] template throughout our analysis. The profile of [O\,{\sc iii}] may be a poor representation of the narrow H$\alpha$ and H$\beta$ emission lines if the NLR is highly stratified in density \citep{Greene2005b}, although, this is usually not the case \citep{Whittle1985,Veilleux1991}. Nevertheless, the nuclear spectra fitting procedure is not particularly sensitive to this election. We tested the use of the [O\,{\sc iii}] narrow core as narrow line shape template, finding that accurate nuclear spectra models can also be obtained in this case. A single Gaussian is sufficient for most of the weaker forbidden lines ([Fe\,{\sc vii}]\,$\lambda 5720$, [Fe\,{\sc vii}]\,$\lambda 6087$, [O\,{\sc i}]\,$\lambda \lambda 6300,6363$ [S\,{\sc ii}]\,$\lambda \lambda 6716, 6731$, [Ar\,{\sc iii}]\,$\lambda 7140$, and [Ar\,{\sc iv}]\,$\lambda 7170$), while two components are usually required for the helium lines (He\,{\sc ii}\,$\lambda 4686$, He\,{\sc i}\,$\lambda 5875$, He\,{\sc i}\,$\lambda 6678$, and He\,{\sc i}\,$\lambda 7065$). With the NLR component thus constrained, we fit broad H$\beta$ and H$\alpha$ with multiple Gaussians, but assign no physical meaning to any of the sub-components; typically four sub-components suffice. We restrict the fit to 4450--7600\,\r{A} in the rest-frame. The lower limit is set to avoid H$\gamma \, \lambda 4340$, which is not included in the MUSE spectral coverage for all sources, while the upper range is limited by the lack of an iron template at longer wavelengths. 

Although the stellar continuum accounts for only a minor fraction of the total flux in the nucleus, we confirm, with the aid of the Bayesian information criterion (BIC; \citealt{Schwarz1978}), that including it in the fit yields statistically improved models for six out of the nine cases ($\Delta$BIC$\,>\,10$).  In one case there is not enough evidence to decide, while in the two cases where the model without the stellar component is favored by the BIC, the differences in the residuals are negligible.  Therefore, for the sake of consistency, all the final fits for the nuclear spectra include the stellar continuum component.  The nuclear spectra and their best-fit models are shown in Figure~\ref{fig:qso_spectra_model_a}.  

\begin{figure*}
\centering
\includegraphics[width=2.0\columnwidth]{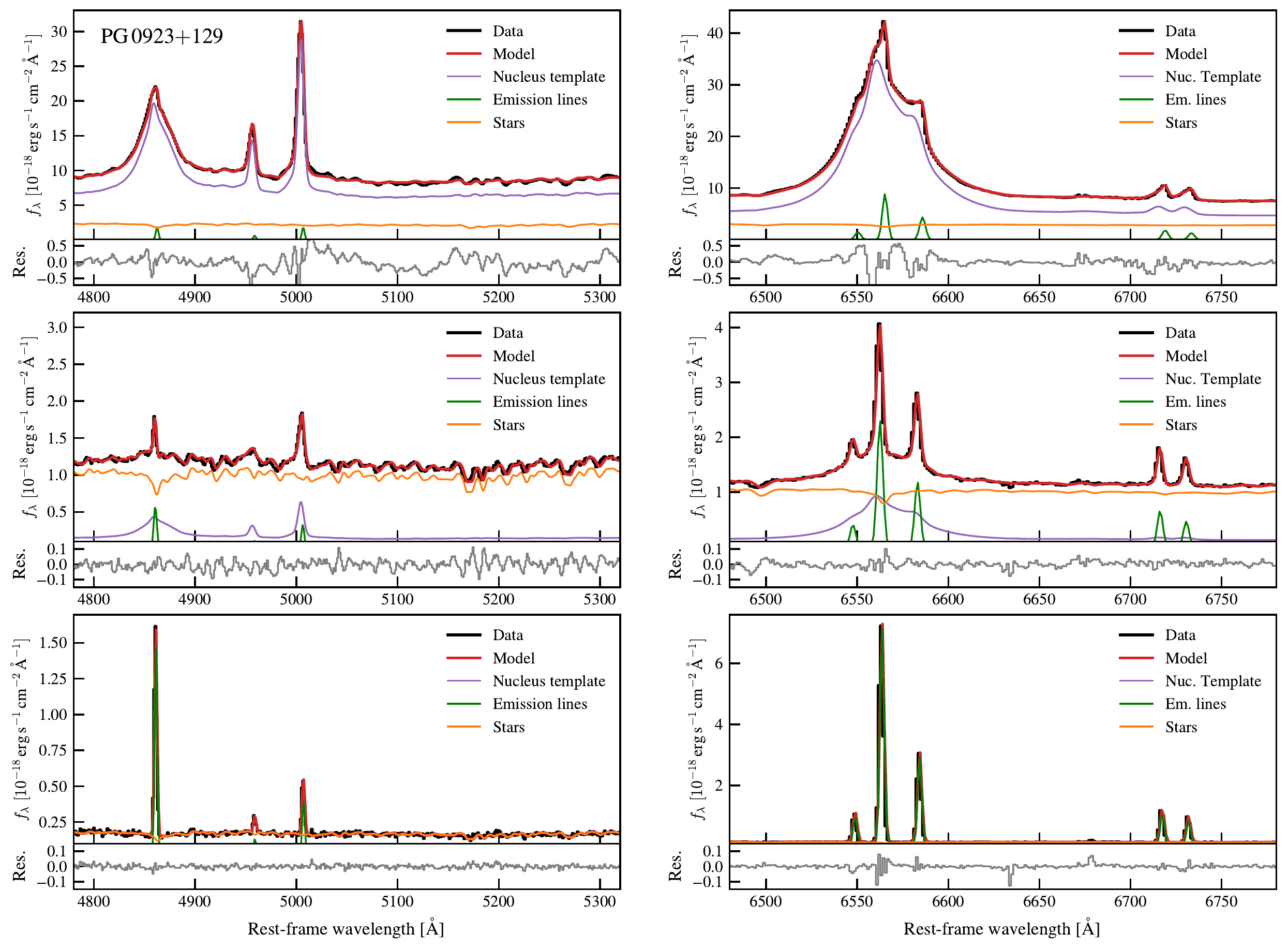}\\
\caption{\label{fig:examplef_fits} Example of fits obtained by our pixel-wise routine for PG\,0923+129. We show three different cases when fitting the spectral windows encompassing the (\textit{left}) H$\beta$+[O\,{\sc iii}] region and (\textit{right}) the H$\alpha$+[N\,{\sc ii}]+[S\,{\sc ii}] region. The top row shows bright blurred nuclear emission; the middle row shows dominant stellar continuum with nuclear scaled contamination; and the bottom row shows a pure  H\,{\sc ii} region emission. The three cases are selected from regions $\sim 1\farcs4$, $3\farcs4$ and $13\farcs8$ away from the quasar peak emission location, respectively. These cases highlight the average quality of the data modeling. In each panel, the bottom subplot gives the residuals of the fit.}
\end{figure*}

\subsection{Characterization of the Host Galaxy Ionized Gas Emission}
\label{sec:Iongas-comp}

We implement a pixel-wise fitting approach to characterize the principal bright diagnostic lines, H$\alpha$, H$\beta$, [O\,{\sc iii}], [N\,{\sc ii}], and [S\,{\sc ii}]. We preferentially preserve the spatial resolution of the observations regardless of the S/N of the lines, in contrast to the Voronoi binning-based strategy adopted when characterizing the stellar component (Section~\ref{sec:Stellar-comp}).

For each pixel, we average the spectrum by considering a box size comparable to the seeing (e.g., \citealt{Swinbank2012a}). The average spectrum is modeled by the sum of a scaled nuclear template, a stellar continuum, plus nebular emission lines. The nuclear template is multiplied by an amplitude free parameter. We assign to each pixel the {\tt pPXF} best-fit stellar continuum model of the corresponding Voronoi cell spectrum (see Section~\ref{sec:Stellar-comp}). Note that due to the strong quasar emission in the center, we do not have a stellar continuum model for the central pixels; for these central pixels, we follow the approach used when deriving the nuclear template, by adopting the best-fit {\tt pPXF} solution of the AGN-free spectra as the stellar continuum template (see Section~\ref{sec:Nuc-comp} for details). The stellar continuum component is simply scaled in amplitude. The emission lines are modeled by using Gaussian functions; in almost all cases only a single component is required. At this step we consider the MUSE instrumental resolution by adding the LSF line width in quadrature within the line model. We use the BIC to determine whether the spectra are well-characterized by our model by comparing the best-fit solution with a simple straight line fit (i.e., no emission present). 

After obtaining the initial set of spectral models, we repeat our procedure using a new set of initial parameters to mitigate the known sensitivity of the least-squares minimization technique to the initial guess. For a given pixel, we take a new set of initial parameters from the best-fit models obtained from the neighboring pixels in the previous run, and from these new solutions we select the model associated with the lowest BIC value. Finally, we use Monte Carlo re-sampling to derive the parameter uncertainties. We measure the average level of the root-mean-square of the spectrum for each pixel from the residuals, add simulated noise to the observed spectrum assuming a normal distribution, and then repeat the fit. We iterate 100 times to obtain a probability distribution for each parameter, from which we estimate the $1\,\sigma$ uncertainties from the 16th and 84th percentiles. 
 
 \begin{figure*}
\centering
\includegraphics[width=1.7\columnwidth]{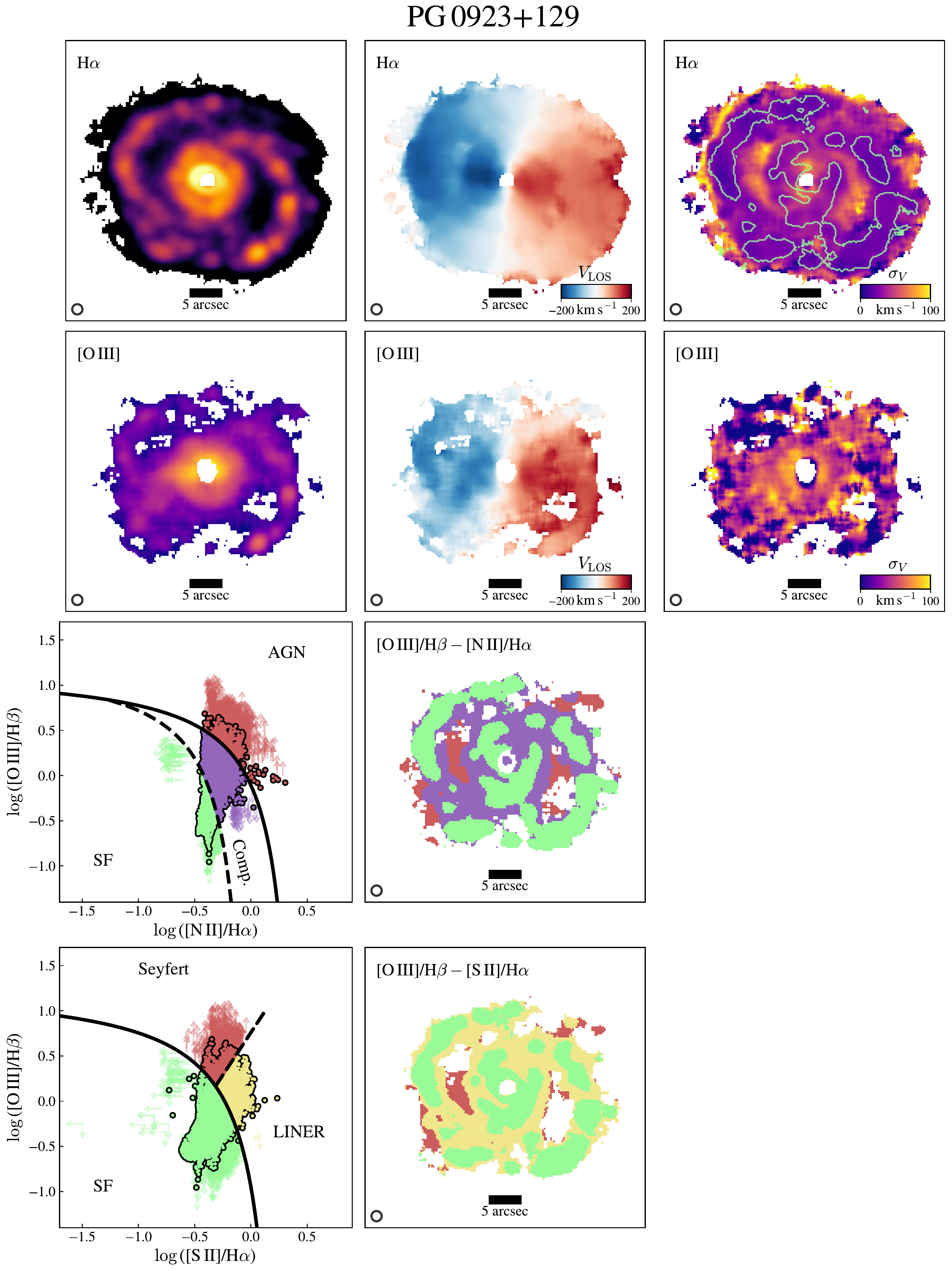}\\
\caption{\label{fig:example_maps} Two-dimensional maps and ionization diagnostic diagrams for PG\,0923+129. The rest of the sample is shown in Appendix~\ref{sec:AppA}. The first two rows show the maps of H$\alpha$ and [O\,{\sc iii}] intensity, LOS velocity, and $\sigma_V$. The third row plots the [O\,{\sc iii}]/H$\beta$-[N\,{\sc ii}]/H$\alpha$ diagram and its spatially resolved two-dimensional distribution. Both panels are color-coded by the spectral classification (AGN, composite and SF) given by the maximum starburst line (solid curve; \citealt{Kewley2001}) and pure star-forming galaxy line (dashed curve; \citealt{Kauffmann2003}). The arrows show the pixels that lack one or two emission-line intensity measurements, but we can still unambiguously classify those regions when considering the line ratio upper/lower limits. The bottom row plots the [O\,{\sc iii}]/H$\beta$-[S\,{\sc ii}]/H$\alpha$ diagram and two-dimensional distribution. In this case, the solid and dashed lines follow the demarcations of \citet{Kewley2001,Kewley2006}. The FWHM of the PSF is shown in the bottom-left corner in each map. In the H$\alpha$ velocity dispersion map, the green contour encloses the pixels classified as SF following the [O\,{\sc iii}]/H$\beta$-[N\,{\sc ii}]/H$\alpha$ diagram, highlighting that these pixels present narrower H$\alpha$ line widths when compared to the others. Nuclear template inaccuracy limits precise emission-line flux measurements near the quasar location when deblending the nuclear emission from the host galaxies.}
\end{figure*}
 
Figure~\ref{fig:examplef_fits} presents example fits of individual pixels of PG\,0923+129, and partly show the models quality that we achieve. Naturally, different correlations among the spectra model parameters arise depending on the target spectra shape. If the blurred nuclear spectra dominates the pixel emission (1st row in Figure~\ref{fig:examplef_fits}), then its amplitude scale parameter tends to be anti-correlated with the H$\alpha$ and H$\beta$ line intensities and line widths. As expected, the nuclear spectrum and stellar continuum scale parameters are anti-correlated when both components significantly contribute to the target spectra emission (2nd row in Figure~\ref{fig:examplef_fits}). When the stellar continuum is significant, its scale amplitude parameter tends to be correlated with the H$\beta$ line intensity because of the presence of the H$\beta$ stellar absorption line. The [O\,{\sc iii}] line intensity tends to be correlated with the [O\,{\sc iii}] line width, but only in regions where the target spectra continuum is faint (3rd row in Figure~\ref{fig:examplef_fits}). Our quasar deblending approach suggests that the blueshifted wing seen in [O\,{\sc iii}] arises from an unresolved region. However, we note that a more sophisticated analysis may be needed to fully constrain the size of the outflow traced by the blueshift seen in [O\,{\sc iii}] (e.g., \citealt{Singha2022}). We emphasize that given the diversity of nuclear spectra shapes observed across our PG quasar sample (Figure~\ref{fig:qso_spectra_model_a}), plus the plethora of combinations between these multiple spectra components digitized in each data cube, the parameter correlations associated with pixel-wise modeling the data must be considered only as a rough description. Figure~\ref{fig:example_maps} gives the two-dimensional maps of the line intensity for H$\alpha$ and [O\,{\sc iii}], their corresponding distributions of line-of-sight (LOS) velocity and velocity dispersion, and the [O\,{\sc iii}]/H$\beta$ versus [N\,{\sc ii}]/H$\alpha$ and [S\,{\sc ii}]/H$\alpha$ diagnostic diagrams and their respective spatial distributions. The velocity dispersions have been corrected for the instrumental spectral resolution. The results for the remaining sources are presented in Appendix~\ref{sec:AppA}.  

\subsection{Classification of the Dominant Ionization Source}
\label{sec:Ion_source}

To classify the dominant ionization source of the nebular emission, we use the line-intensity ratio diagnostic diagrams [O\,{\sc iii}]\,$\lambda 5007$/H$\beta$ versus [N\,{\sc ii}]\,$\lambda 6584$/H$\alpha$ and [O\,{\sc iii}]\,$\lambda 5007$/H$\beta$ versus [S\,{\sc ii}]\,$\lambda \lambda 6716, 6731$/H$\alpha$ (\citealt{Baldwin1981}; hereinafter BPT). The two diagrams provide different levels of differentiation in terms of the ionization mechanism.  The [O\,{\sc iii}]/H$\beta$-[N\,{\sc ii}]/H$\alpha$ diagram designates three broad classes (Figure~\ref{fig:example_maps}): objects below the dashed line are powered by star formation \citep{Kauffmann2003}; objects above the ``maximum starburst'' solid line are powered by AGNs \citep{Kewley2001}; and those in between these two boundaries have a composite source of ionization (mixture of young stars and AGNs).  Classifications based on the [O\,{\sc iii}]/H$\beta$-[S\,{\sc ii}]/H$\alpha$ diagram use the demarcations given by \citet{Kewley2001,Kewley2006}, which distinguish star-forming objects from Seyferts and low-ionization nuclear emission-line regions (LINERs). The properties of low-luminosity AGNs, including the physical distinction between high-ionization (Seyfert) and low-ionization (LINER) sources have been extensively reviewed by \citet{Ho2008b,Ho2009b} and will not be repeated here. We assign a classification to all pixels whose emission lines were detected at S/N\,$>$\,3, and to those whose ionization source can be classified unambiguously even with just emission-line flux limits. 
For the former pixels, all emission-line fluxes are corrected for internal reddening based on the observed Balmer decrement of the narrow lines, assuming an intrinsic value of H$\alpha$/H$\beta$ = 2.86 for the star-forming pixels and H$\alpha$/H$\beta$ = 3.1 for the others \citep{Osterbrock2006}, assuming the extinction curve of \citet{Cardelli1989}.

\section{Results}
\label{sec:Results}

\subsection{Narrow-line Region}
\label{sec:NLR_all}

Often characterized by its extension and geometry, the NLR reflects how the ionizing radiation from the central AGN permeates through the ISM of the galaxy host. It also traces the deposition of kinetic energy and momentum when outflows are present (e.g., \citealt{Greene2012,Husemann2013,Husemann2016,Husemann2019a, Storchi2018,SunAGN2018})  Figures~\ref{fig:example_maps} and \ref{fig:all_maps} reveal that the PG quasar hosts have vast ISM zones in which the emission is consistent with being powered by the AGN, allowing us to characterize in detail their NLR.\footnote{Similar to other IFU studies (e.g., \citealt{Husemann2016,Husemann2019b}), our quasar deblending technique already separates the unresolved nuclear NLR emission from the resolved NLR emission coming from the large-scale ISM of the host galaxy.}

\subsubsection{NLR Size-Luminosity Correlation}
\label{sec:NLR_LR}

\begin{figure}
\centering
\includegraphics[width=1.0\columnwidth]{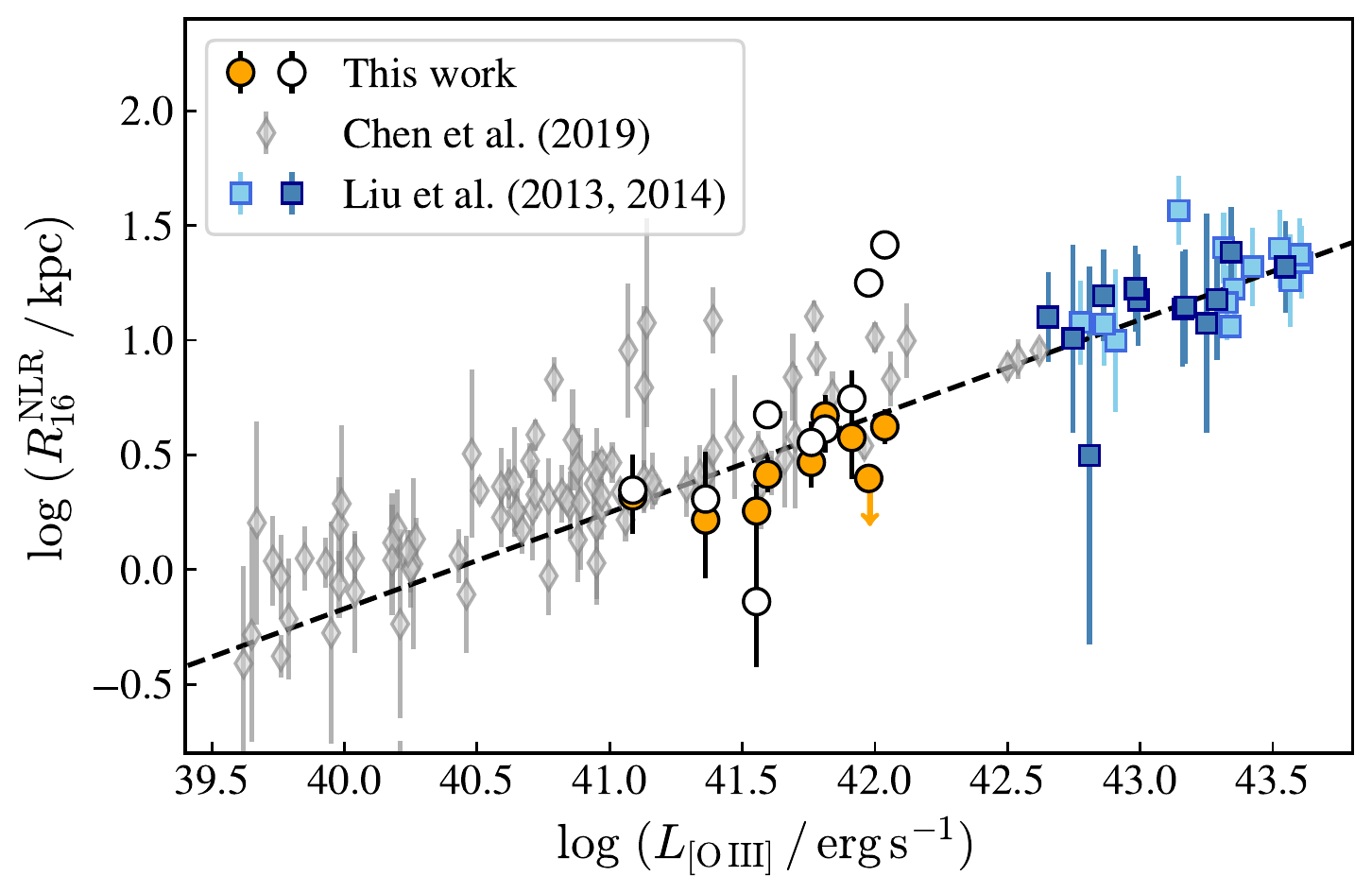}\\
\caption{\label{fig:NLR_LS} The [O\,{\sc iii}] emission-line luminosity as a function of the NLR radial size for the PG quasar hosts. The orange circles represent the best-fit isophote radius at an [O\,{\sc iii}] surface brightness of 10$^{-16}$\,erg\,s$^{-1}$\,cm$^{-2}$\,arcsec$^{-2}$\,(1+$z$)$^{-4}$. The open circles represent the maximum NLR size given by the most distant gaseous region with an [O\,{\sc iii}] surface brightness value higher than the 10$^{-16}$\,erg\,s$^{-1}$\,cm$^{-2}$\,arcsec$^{-2}$\,(1+$z$)$^{-4}$ limit. The grey diamonds show the MaNGA survey data \citep{Chen2019}. The sky-blue and blue squares correspond to the $z \approx 0.5$ type~2 and type~1 quasar IFU data presented in \citet{Liu2013,Liu2014} and re-analyzed by \citet{Chen2019}. The dashed line corresponds to the best linear fit (slope\,$\sim 0.42$) computed by \citet{Chen2019}. The arrow shows our upper limit estimate for the NLR isophote size in PG\,0050+124. The PG quasar host galaxies follow the NLR luminosity-size relation.}
\end{figure}

The correlation between the NLR size and the AGN [O\,{\sc iii}] luminosity ($L_{[{\rm O\,III}]}$) is usually employed to constrain the properties of the gas that is photoionized by the central source (e.g., \citealt{Bennert2002,Liu2013,Liu2014,Hainline2014,Dempsey2018}). We compute extinction-corrected $L_{[{\rm O\,III}]}$ by summing all S/N\,$>3$ line emission from pixels classified as ``Seyfert''  according to the [O\,{\sc iii}]/H$\beta$-[S\,{\sc ii}]/H$\alpha$ diagram. We confirm that $L_{[{\rm O\,III}]}$ scales with the AGN bolometric luminosity, which we estimate from $L_{5100}$, following the trend reported by \citet{Pennell2017} within a scatter of 0.3\,dex. While many different strategies for measuring NLR radial sizes have been used in the literature (see Section~3.1 of \citealt{Husemann2019b}), we focus on two common procedures; (1) we follow \citet{Chen2019} and fit the NLR radial profiles using S\'ersic functions \citep{Sersic1963} to characterize the NLR radius by the isophote at an intrinsic [O\,{\sc iii}] surface brightness of 10$^{-16}$\,erg\,s$^{-1}$\,cm$^{-2}$\,arcsec$^{-2}$\,(1+$z$)$^{-4}$ and measured from the Seyfert-classified pixels ($R^{\rm NLR}_{\rm 16, fit}$); and (2) we estimate the maximum NLR radial size ($R^{\rm NLR}_{\rm 16, max}$) by computing the deprojected location of the most distant AGN-ionized region emitting above the 10$^{-16}$\,erg\,s$^{-1}$\,cm$^{-2}$\,arcsec$^{-2}$\,(1+$z$)$^{-4}$ [O\,{\sc iii}] surface brightness threshold (e.g., \citealt{Husemann2022}). In both cases we account for the effect of cosmic dimming, enabling the comparison of AGN hosts observed at different redshifts \citep{Liu2013}. We account for inclination correction by measuring the minor-to-major axis ratio of the host galaxy from the white-light image using \textsc{photutils}. We report those estimates in Table~\ref{tab:derived_quantities}. Figure~\ref{fig:NLR_LS} compares our $R^{\rm NLR}_{\rm 16, fit}$ measurements of the PG quasars hosts with the large sample of 152 Seyfert galaxies (both type~1 and type~2, but mostly type~2) studied by \citet{Chen2019} using the Mapping Nearby Galaxies at Apache Point Observatory (MaNGA) survey \citep{Bundy2015}. Also included are luminous type~1 and type~2 quasars at $z\approx 0.5$ from \cite{Liu2013,Liu2014}. With $R^{\rm NLR}_{\rm 16, fit} \approx 1.6-4.7$\,kpc (Table~\ref{tab:derived_quantities}), the PG quasars follow the best-fit NLR size-$L_{[{\rm O\,III}]}$ relation previously reported by \citet{Chen2019}. We only give an upper limit for the NLR size of PG\,0050+124 because we cannot detect extended [O\,{\sc iii}] emission from its central $R \lesssim2.5\,$kpc zone (see Appendix~\ref{sec:AppA}). 

Interestingly, among all the PG quasar hosts, the zones where the AGN mainly contributes to the ionization state of the ISM coincide with the regions where H$\alpha$ emission is faint and the ionized gas follows the global rotation pattern of the galaxy (Figures~\ref{fig:example_maps} and \ref{fig:all_maps}), suggesting that this gas was uplifted from the disk due to local secular processes before being ionized by the AGN \citep{Husemann2019b}, or simply that the AGN-ionized gas emission is outshined by the H\,{\sc ii} region ionizing photons where the star formation activity takes place. Geometrical effects may play a substantial role when determining the sizes of the NLR from simple profile fits. To highlight this, in Figure~\ref{fig:NLR_LS} we show in open circles the $R^{\rm NLR}_{\rm 16, max}$ values for the PG quasars, and we compare with the $R^{\rm NLR}_{\rm 16, fit}$ values (orange circles). Both estimates are in agreement within the uncertainties for most of the cases, but three outliers can be clearly seen. The PG quasar host galaxies with consistent $R^{\rm NLR}_{\rm 16, fit}$ and $R^{\rm NLR}_{\rm 16, max}$ estimates tend to show smooth [O\,{\sc iii}] surface brightness maps, and with the brighter zones being detected closer to the quasar location. In one of the outliers, PG\,0923+129, $R^{\rm NLR}_{\rm 16, max} \approx 0.7$\,kpc is significantly smaller than $R^{\rm NLR}_{\rm 16, fit} \approx 1.8$\,kpc.  Careful scrutiny reveals that $R^{\rm NLR}_{\rm 16, fit}$ is given by an extrapolation of the radial profile in a location where this host galaxy contains an inner star-forming ring seen in H$\alpha$ (Figure~\ref{fig:example_maps}) and in CO \citep{Molina2021}, suggesting that the smaller $R^{\rm NLR}_{\rm 16, max}$ value measured for this system may be produced by the AGN-ionized gas emission being outshone by the ionizing photons coming from the underlying star formation activity in that local ISM sub-structure. Another possibility could be that the dust on the star-forming ring might be attenuating the number of AGN ionizing photons that can permeate to larger radii. In the other two cases we have the opposite evidence, where $R^{\rm NLR}_{\rm 16, max}$ values are significantly higher than the $R^{\rm NLR}_{\rm 16, fit}$ estimates. One case corresponds to PG\,1426+015, a host galaxy that has two tidal features that are being ionized by the AGN, with the most distant feature $\sim 26$\,kpc away from the nucleus presenting [O\,{\sc iii}] emission above the surface brightness threshold adopted to define the NLR size. Similarly, PG\,0050+124 also displays a set of distant ($\sim 18$\,kpc) ionized clouds emitting [O\,{\sc iii}] above the surface brightness threshold, which is surprising given that we only found a $R^{\rm NLR}_{\rm 16, fit}$ upper limit value for this target. With $\log\,$[N\,{\sc ii}]/H$\alpha \approx -0.5$ and $\log\,$[O\,{\sc iii}]/H$\beta \approx 1.0$, the gas composition appears to be metal-poor, not unlike the situation found in HE\,1353$-$1917 \citep{Husemann2019b}. We note that some of the systems analyzed by \citet{Chen2019} also have NLR sizes significantly above their best-fit relation (Figure~\ref{fig:NLR_LS}). They rejected merger pairs when designing their sample. Nearly half of their systems with $R^{\rm NLR}_{\rm 16, fit} \gtrsim 5$\,kpc tend to be $z \approx 0.10-0.15$ galaxies suggesting that the MaNGA observation's spatial resolution ($\sim 2\farcs5$) may be a concern.

Overall, our findings suggest that, as expected, the surface brightness of the NLR depends of the central AGN radiation power, but the NLR extension is mainly determined by geometrical factors. The exemplary cases are PG\,0050+124 and PG\,1426+015 where the NLR emission is detected abnormally bright at the outskirts of the host galaxies (Figure~\ref{fig:example_maps}). In the PG quasars, the NLR extension is probably limited by the availability of gas that can be ionized by the AGN radiation (matter-bounded case), and not by the lack of AGN ionizing photons at large galactocentric radius (ionization-bounded case). This finding is in line with the reports of other examples of distant AGN-photoionized gas debris and filaments, as well as large-scale gas outflows located beyond the main disk of the host galaxy disk (e.g., \citealt{Greene2012,Storchi2018,Villar2018,Tubin2021}). 

\begin{figure*}
\centering
\includegraphics[width=1.8\columnwidth]{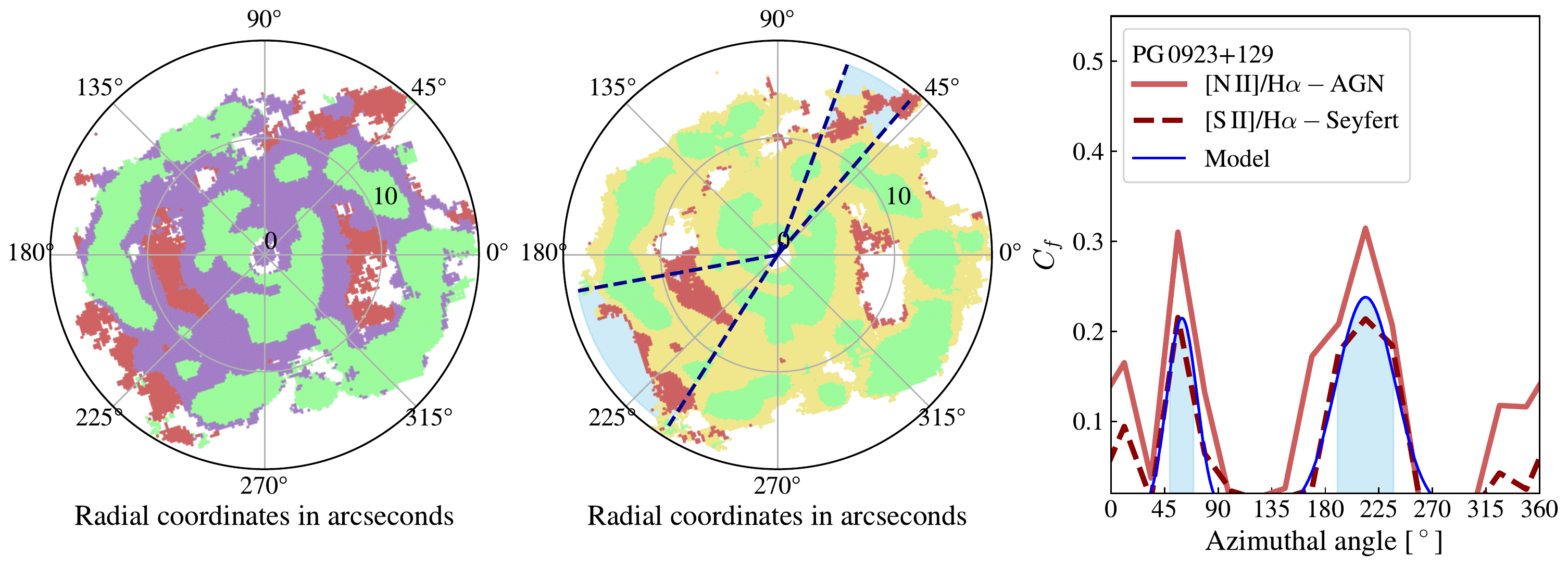}\\
\caption{\label{fig:polar_maps} The host galaxy of PG\,0923+129 seen in polar projection. Line-ratio diagnostic maps for (left) [N\,{\sc ii}]/H$\alpha$ and (middle) [S\,{\sc ii}]/H$\alpha$, color-coded following Figure~\ref{fig:example_maps}. The radial coordinates are represented by the grey circles. If detected, the ionization cones are represented by blue shaded regions and delimited by the blue dashed lines in the [S\,{\sc ii}]/H$\alpha$ maps. The right plot gives the covering fraction of the pixels classified as AGN according to [N\,{\sc ii}]/H$\alpha$ and as Seyfert according to [S\,{\sc ii}]/H$\alpha$, as a function of the projected azimuthal angle. The blue solid lines show the best fits to the $C_f$ values for the [S\,{\sc ii}]/H$\alpha$ map, if a clear excess is detected. The blue shaded zones under the Gaussian models ($\pm 1\,\sigma$) represent the same azimuthal angle coverage shown in the [S\,{\sc ii}]/H$\alpha$ map. The rest of the sample can be found in Appendix~\ref{sec:AppB}.}
\end{figure*}

\begin{figure}
\centering
\includegraphics[width=0.7\columnwidth]{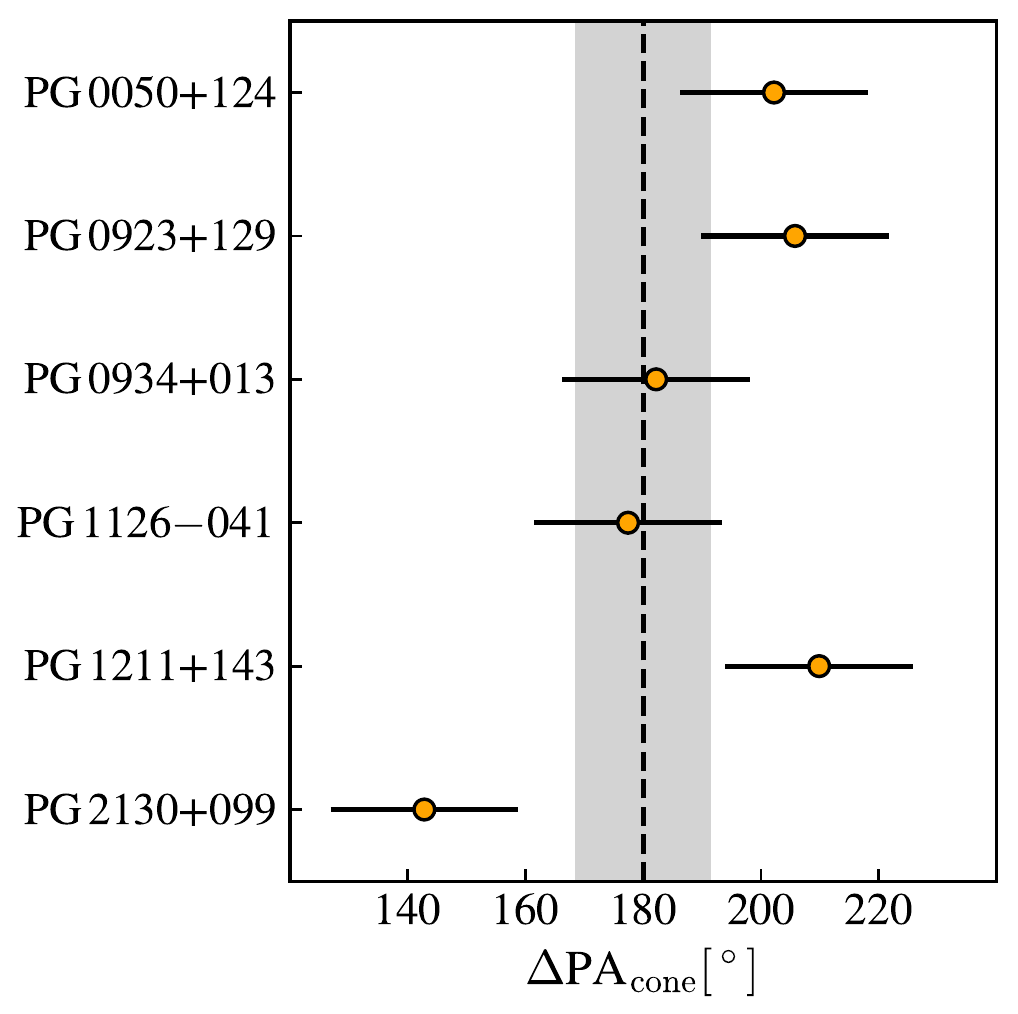}\\
\caption{\label{fig:cone_cdiff} Azimuthal angular difference between the ionization cone side position angles for the PG quasar host galaxies with two-sided ionization cones. The dashed line indicates the ideal 180$^\circ$ angle difference between both sides. The shaded zone represents the azimuthal angular size bin ($22.5^\circ$) used to compute the $C_f$ values from the two-dimensional maps; for each ionization cone PA value we assume an uncertainty equal to half the bin size. The data are clustered around the expected azimuthal angle difference, consistent with the bipolar geometry of an ionization cone.}
\end{figure}

\subsubsection{NLR Geometry}
\label{sec:NLR_geom}

To explore the NLR geometry we analyze the ionization energy source maps in polar projection (Figure~\ref{fig:polar_maps}). Aiming to identify possible bipolar ionization cones, we first bin the maps in 16 azimuthal bins, and for each individual bin and we compute the covering fraction of the AGN-classified pixels as,

\begin{equation}
\label{eq:C_f}
C_f = N_{\rm pix}^{\rm AGN} /  N_{\rm pix}^{\rm tot},
\end{equation}

\noindent where $N_{\rm pix}^{\rm AGN}$ is the number of pixels classified as AGN-like according to the [N\,{\sc ii}]/H$\alpha$ map and as Seyfert-like according to the [S\,{\sc ii}]/H$\alpha$ map, and $N_{\rm pix}^{\rm tot}$ is the number of pixels in which we detect any emission that can be classified in the BPT maps (Section~\ref{sec:Ion_source}). Figure~\ref{fig:polar_maps} presents the trends of $C_f$ with respect to the azimuthal angle for both BPT classification schemes for example object PG\,0923+129, while the rest of our sample is presented in Appendix~\ref{sec:AppB}. For most of the systems we find that $C_f$ peaks at different azimuthal angle values in both BPT maps, but we caution that the presence of star-forming sub-structures (e.g., star-forming ring in PG\,0923+129 or spiral arms in PG\,0050+124) may outshine the AGN-induced emission and/or shape the NLR geometry. We focus on the $C_f$ values estimated from the [S\,{\sc ii}]/H$\alpha$ map, as this classification scheme better separates between emission produced by AGN photoionization and shocks.\footnote{As reviewed in \citet{Ho2008b}, LINER emission far from the central region of galaxy can be powered by a variety of sources, including shocks. This should not be conflated with the proposition that LINERs, especially in low-luminosity AGNs, bear no relation to BH accretion. As this study concerns highly accreting, luminous quasars, it is reasonable to attribute LINER-like emission to non-nuclear sources of excitation, such as shocks, and only attribute to AGN photoionization those regions classified as Seyfert-like.} Six of the objects show an increase of $C_f$ at particular bins of azimuthal angle, suggesting that the AGNs ionize gas along preferred position angles. For the remaining sources we do not find any significant trend with respect to azimuthal angle. One of these is PG$\,$1426+015, a merger for which we cannot determine any NLR geometry given its complex morphology. For the two remaining systems PG\,1011$-$040 and PG\,1244+026, we speculate that the observed round geometry of its NLR may be consistent with the source being viewed through a LOS close to the direction of the ionization cone (pole-on view; e.g., \citealt{Storchi2018}). Observing larger samples of type~1 and type~2 AGNs would help to understand whether this possibility is consistent with the expectations implied by the AGN unification model from a statistical point of view.

We characterize the sources presenting preferred AGN ionization direction by fitting the $C_f$ values using up to two Gaussians. During this modeling we account for the azimuthal binning implemented to estimate the $C_f$ values, and we also add a constant free parameter value that helps to deal with cases where few Seyfert-classified pixels are detected in all directions (e.g., PG\,1126$-$041). We assume that the centroids of the Gaussian trace the position angles of the ionization cone sides, while the width of the Gaussian represents the opening angle of the cone (likely to be lower limits due to surface brightness bias). Hence, each Gaussian is assumed to represent one side of the ionization cone, and a one-sided ionization cone is well-represented by only one Gaussian component. The inferred location of the projected ionization cones is overlaid on the [S\,{\sc ii}]/H$\alpha$ map in Figures~\ref{fig:polar_maps} and \ref{fig:polar_maps_app}. We find two-sided ionization cones in PG\,0050+124, PG\,0934+013, PG\,1126$-$041, and PG\,1211$+$143, and more subtle evidence is suggested in PG\,0923+129 and PG\,2130+099. To determine if these two-sided ionization cones are consistent with a biconical geometry, in Figure~\ref{fig:cone_cdiff} we show the difference between the ionization cone sides position angles ($\Delta {\rm PA}_{\rm cone}$) for these six quasar hosts.  We find a mean position angle difference of $\sim 187^\circ$, in close agreement with the expected difference of $180^\circ$ for a bipolar cone geometry. We measure a scatter of $\sim 23^\circ$, which is comparable to the azimuthal bin size ($22.5^\circ$) employed to tessellate the host galaxy maps. Admittedly, our NLR geometry analysis is limited to the systems where the ionization cones are misaligned with respect to the observation line-of-sight. While $C_f$ is conveniently defined to be immune to the host galaxy inclination angle, it may not be useful to study the NLR geometry in highly inclined objects.

\subsection{Ionized Gas Outflows}
\label{sec:outflows}

The association of disturbed ionized gas kinematics with ionization cones has been interpreted as evidence for ionized gas outflows in active galaxies (e.g., \citealt{SunAGN2017,Kang2018}). While ionization cones can extend up to several kpc within the host galaxy or beyond, the ionized gas outflows tend to be primarily detected on smaller scales (e.g., \citealt{Karouzos2016,Fischer2018,Husemann2019b}). Our MUSE maps of PG quasar host galaxies offer us the opportunity to explore the possible contribution of outflows to AGN feedback. Outflows can prevent or delay star formation if they succeed in evacuating or relocating a significant fraction of the ISM \citep{Fabian2012}. Conversely, outflows can promote star formation through gas compression by shocks \citep{Cresci2015,Maiolino2017}. 

\begin{figure}
\centering
\includegraphics[width=1.0\columnwidth]{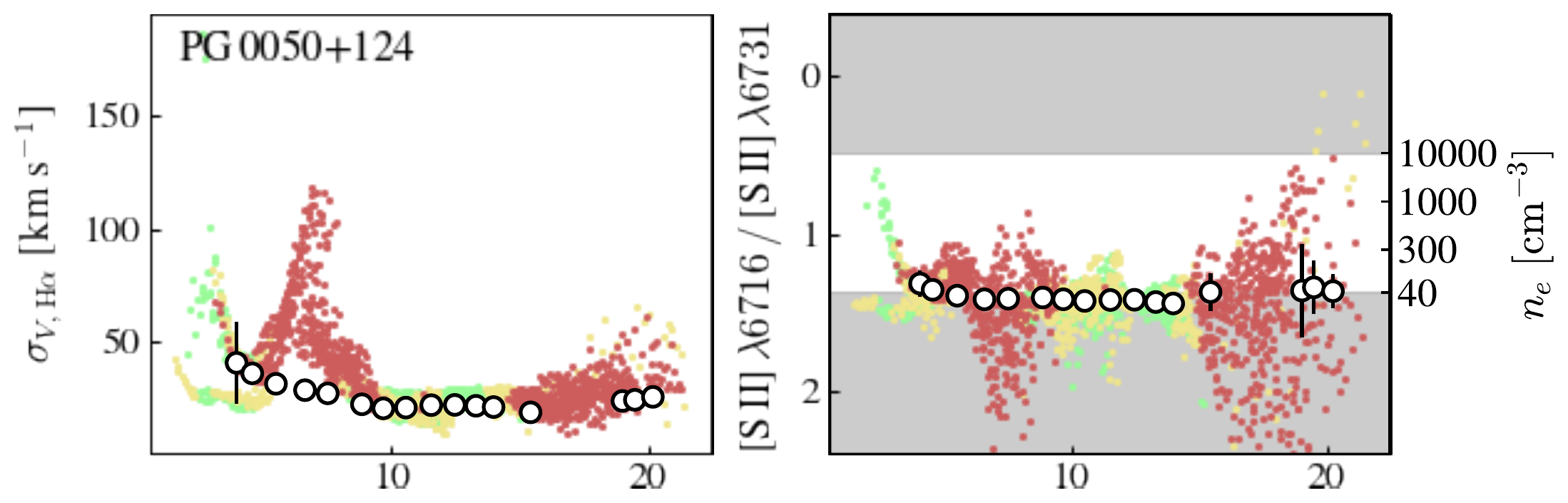}\\
\includegraphics[width=1.0\columnwidth]{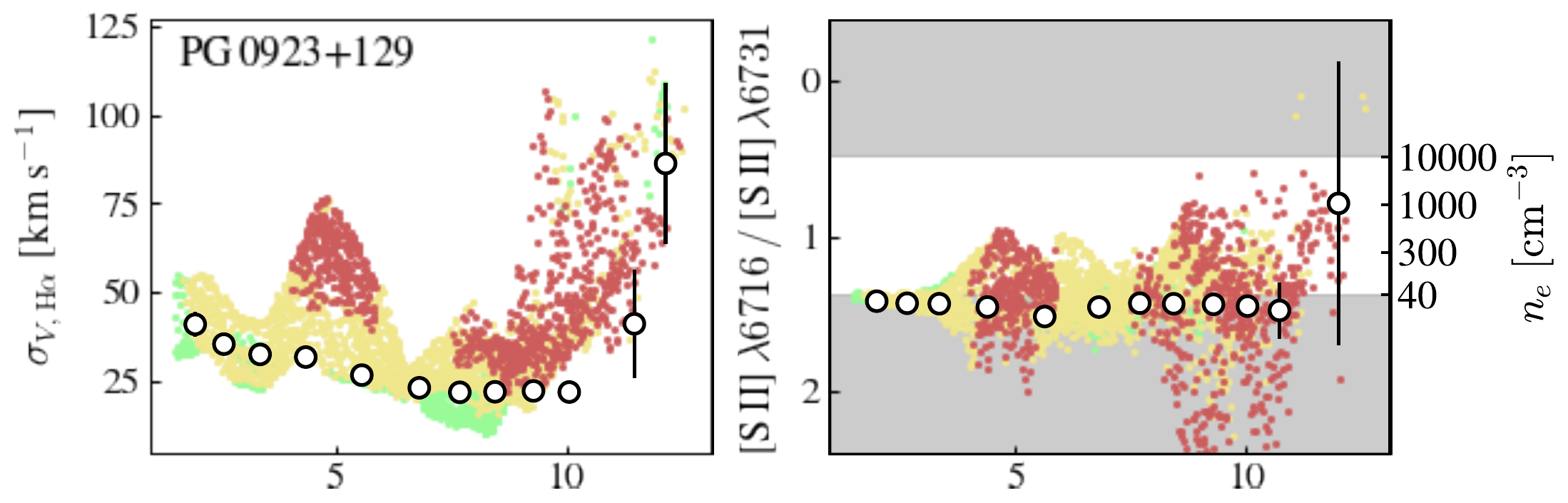}\\
\includegraphics[width=1.0\columnwidth]{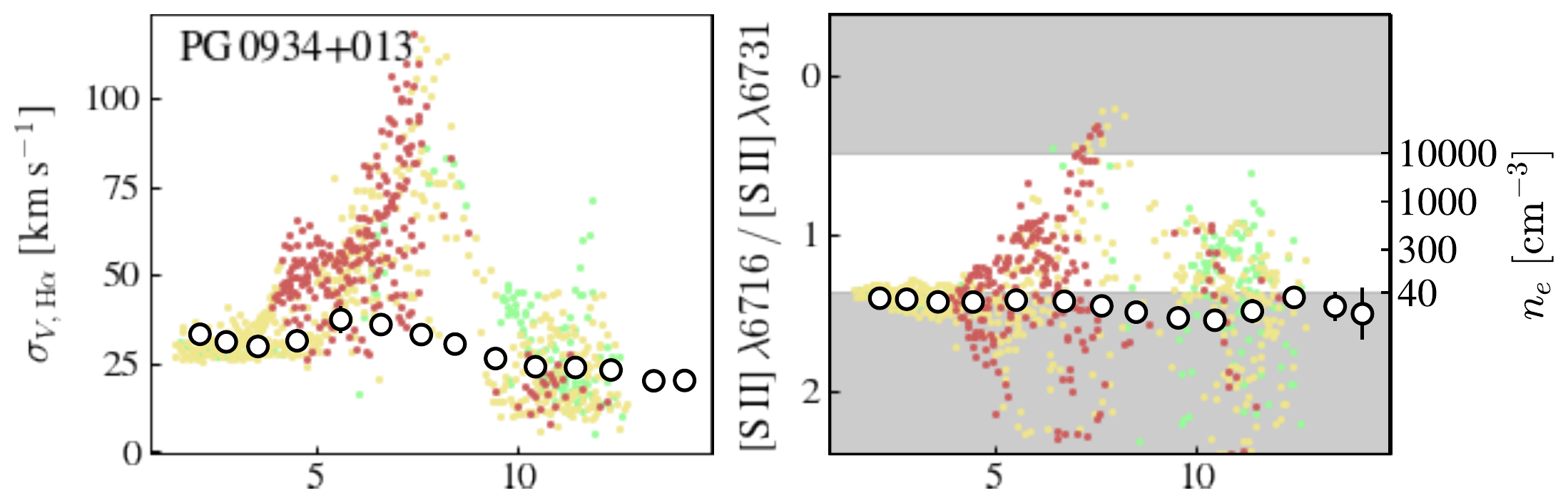}\\
\includegraphics[width=1.0\columnwidth]{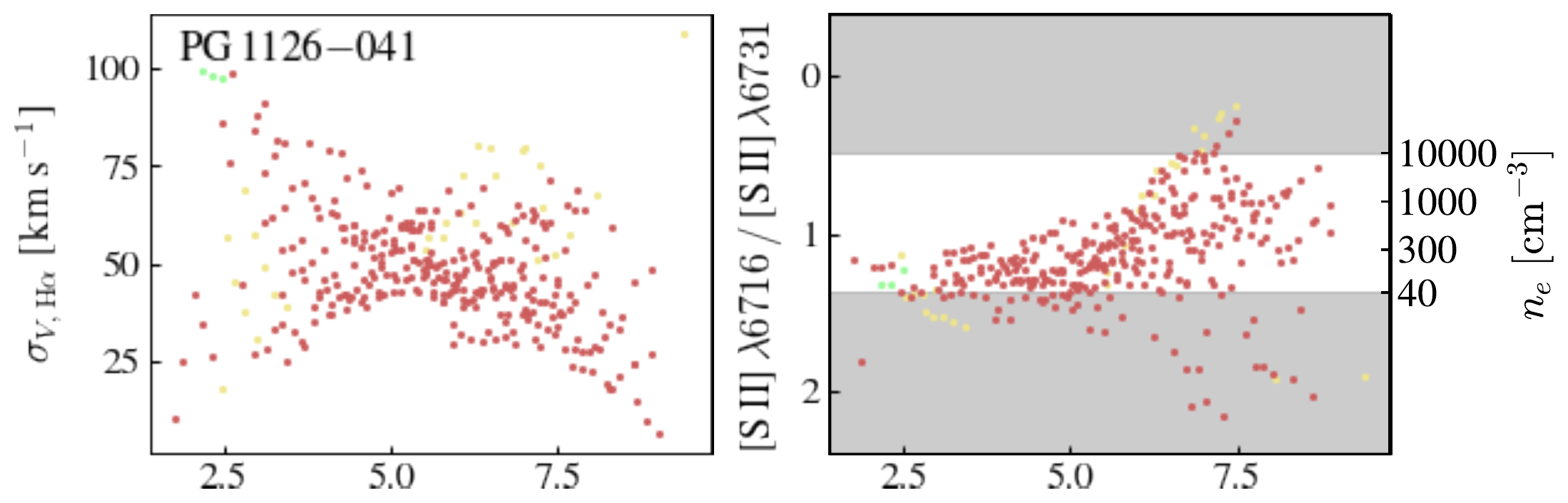}\\
\includegraphics[width=1.0\columnwidth]{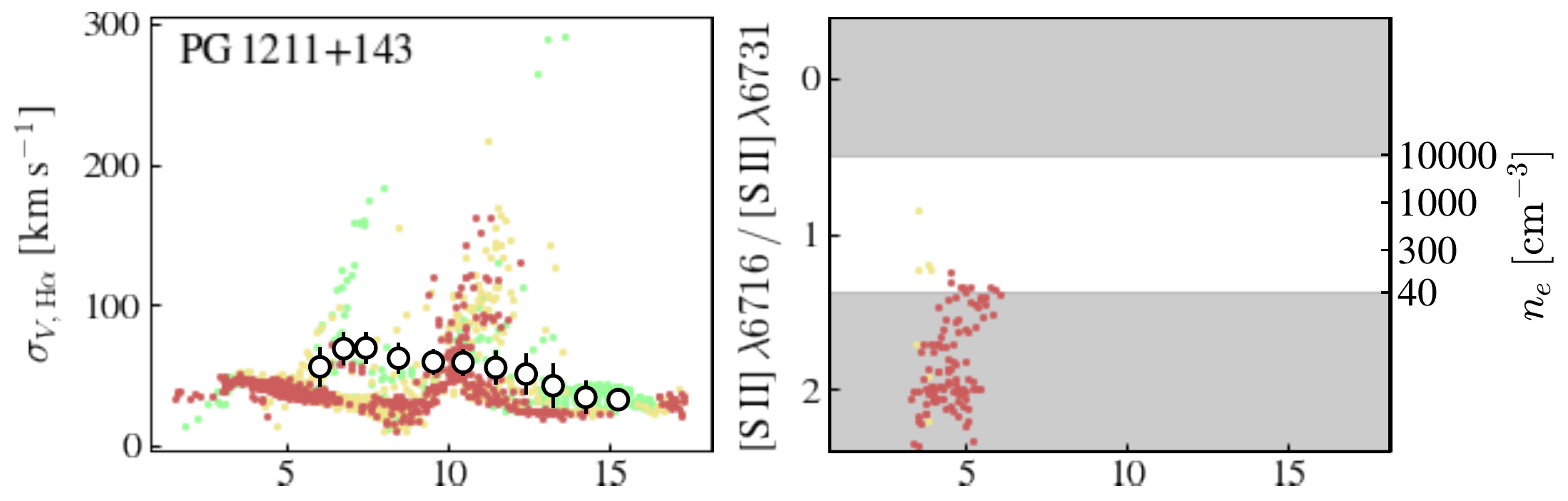}\\
\includegraphics[width=1.0\columnwidth]{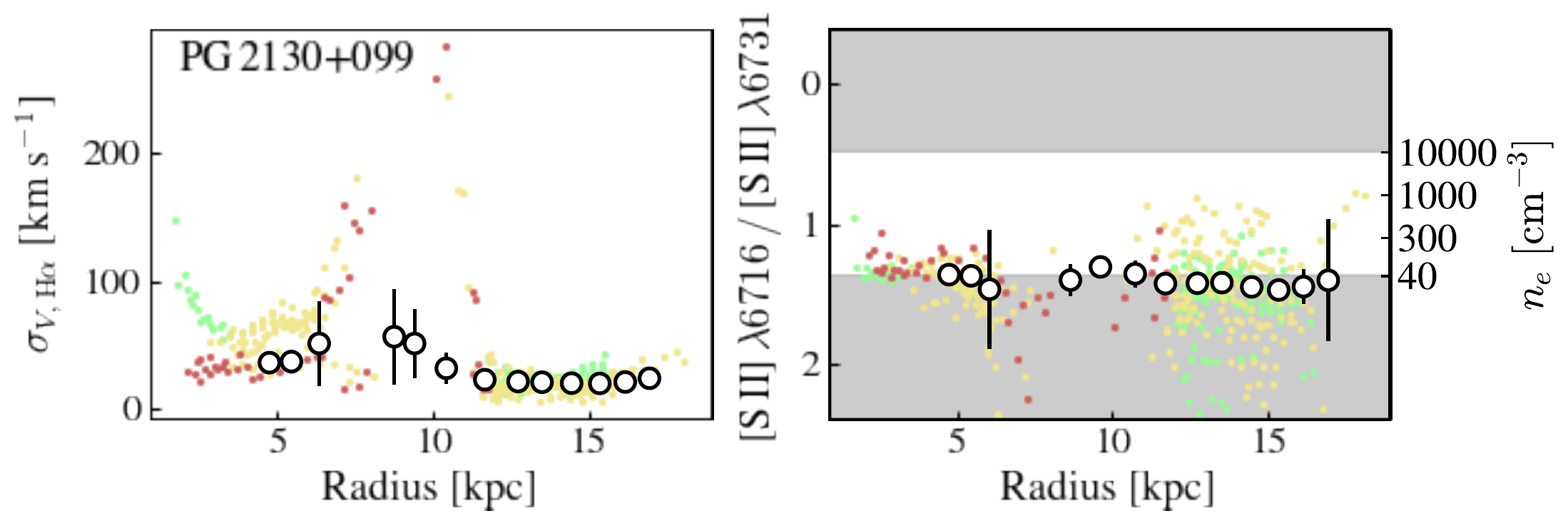}\\
\caption{\label{fig:cone_kinprofs} Pixel-wise H$\alpha$ velocity dispersion (left) and [S\,{\sc ii}] emission-line intensity ratio (right) across the axis of the detected ionization cones. In both panels the data are color-coded following the [S\,{\sc ii}]/H$\alpha$ maps of Figure~\ref{fig:example_maps}, and the open black circles represent azimuthal average values computed from the pixels labeled as star-forming (according to the [N\,{\sc ii}]/H$\alpha$ classification scheme), if any. In the right panels, the unshaded region denotes the range for which the [S\,{\sc ii}]\,$\lambda 6716$/[S\,{\sc ii}]\,$\lambda 6731$ ratio is sensitive to $n_e$ ($\sim 40-10^4$\,cm$^{-3}$), with the corresponding $n_e$ scale given in the right axis, following the conversion of \citet{Proxauf2014}. All objects exhibit regions with enhanced $\sigma_{V, \rm H\alpha}$, although $n_e$ increases only in PG\,0934+013 and PG\,1126$-$041.}
\end{figure}

\subsubsection{Local Observational Signatures}
\label{sec:outflows_obsig}

Visual inspection of the host galaxy maps (Figures~\ref{fig:example_maps} and \ref{fig:all_maps}) suggests that the lower H$\alpha$ velocity dispersions ($\sigma_{V, \rm H\alpha}$) tend to be measured in zones where the star-forming regions mainly contribute to the ionization state of the ISM (pixels enclosed by green contours in the $\sigma_{V, \rm H\alpha}$ maps). Interestingly, in some sources (e.g., PG\,0050+124 and PG\,0934+013) the pixels with higher $\sigma_{V, \rm H\alpha}$ are located in zones that overlap with the direction of the ionization cone. This is not the case when observing the [O\,{\sc iii}] velocity dispersions ($\sigma_{V, \rm [OIII]}$), where higher values do not spatially coincide with the high $\sigma_{V, \rm H\alpha}$ estimates in some cases (e.g., PG\,1126$-$041). However, the [O\,{\sc iii}] emission line is observed in the bluer region of the MUSE spectra coverage implying that the low $\sigma_{V, \rm [OIII]}$ estimates are less reliable when compared to the $\sigma_{V, \rm H\alpha}$ values due to the MUSE LSF variation with wavelength, i.e., the [O\,{\sc iii}] lines are observed at coarser spectral resolution and subject to a more uncertain line width deconvolution correction. Hence, in this study we focus on studying the apparent connection between the H$\alpha$ velocity dispersion with respect to the ionization state of the gas. We also inspect the gas electron density ($n_e$) variation across the ionization cone direction, as this quantity may reflect the compression of the gas by shocks \citep{Osterbrock2006}. To this end, we take advantage of the respective sensitivity of both BPT maps, using the [S\,{\sc ii}]/H$\alpha$ map to isolate the effect of AGN ionization and the [N\,{\sc ii}]/H$\alpha$ map to highlight the star-forming regions. 

Figure~\ref{fig:cone_kinprofs} examines the deprojected radial variation of $\sigma_{V, \rm H\alpha}$ and [S\,{\sc ii}]\,$\lambda 6716$/[S\,{\sc ii}]\,$\lambda 6731$ across the six detected ionization cones. We convert the ratio of the [S\,{\sc ii}] doublet to $n_e$ adopting an electron temperature of $10^4$\,K following Eq.~3 of \citet{Proxauf2014}. Consequently, we assume that the [S\,{\sc ii}] emission only provides a reliable estimate of $n_e$ within the $40-10^4$\,cm$^{-3}$ range. However, we caution that the $n_e$ values correspond to the line-of-sight integrated densities instead of local ISM estimates. Across all six ionization cones we find an increase in gas velocity dispersion up to $\sigma_{V, \rm H\alpha} \approx 100-200$\,km\,s$^{-1}$ at specific radii.  The only exception is PG\,1126$-$041, whose $\sigma_{V, \rm H\alpha}$ profile continuously increases as a function of radius. In all the cases, the regions associated with an increase of $\sigma_{V, \rm H\alpha}$ tend to be ionized by the AGN, according to the [S\,{\sc ii}]/H$\alpha$ diagram classification. In contrast, the pixels classified as star-forming from the [N\,{\sc ii}]/H$\alpha$ map have relatively smooth radial trends in velocity dispersion in the range $\sigma_{V, \rm H\alpha} \approx 20-45$\,km\,s$^{-1}$ (we omit PG\,1126$-$041, which has too few star-forming pixels), $\sim 5$ times lower than those in the AGN-classified locations. We find no clear trends in terms of [S\,{\sc ii}]\,$\lambda 6716$/[S\,{\sc ii}]\,$\lambda 6731$. PG\,0934+013 is the only system in which an increase in $\sigma_{V, \rm H\alpha}$ spatially coincides with a decrease in [S\,{\sc ii}]\,$\lambda 6716$/[S\,{\sc ii}]\,$\lambda 6731$ (increase of $n_e$); the most turbulent, AGN-dominated regions have densities $\gtrsim 200$ times higher than those in the star-forming locations ($n_e \lesssim 40$\,cm$^{-3}$). That $\sigma_{V, \rm H\alpha}$ only increases at specific radii within the host suggests that the axis of the ionization cone is not pointed along the gas disk, but instead is slightly offset, such that it affects the ISM at large vertical scale heights where diffuse ionized gas lies.  Such a disk-outflow geometric configuration may imply that the outflows are unable to significantly affect the gas located in the disk midplane where the star-forming complexes lie. The outflows do not dominate the ionization, so that the flux-weighted emission coming from co-spatial H\,{\sc ii} regions mainly reflects the star formation process, and hence the Seyfert-classified regions are mainly detected where the H$\alpha$ emission generated by the star formation activity is faint (see Figures~\ref{fig:example_maps} and \ref{fig:all_maps}). Instead of a homogeneously filled cone aligned with respect to the disk midplane, the outflow geometry may be more akin to that of an inclined expanding shell, which is capable of producing only a local enhancement in gas density and turbulence at the shock front interface with the gas clouds (e.g., expanding hot gas cocoon in Figure~5 of \citealt{Husemann2019b}).  We caution that we cannot rule out the possibility that we are missing some inner faint emission and kinematic fingerprint induced by the outflow because the weak signal is swamped by brighter emission from star-forming complexes across the same LOS (e.g., PG\,0923+129 and PG\,0934+013). Our ability to detect kinematically subtle signatures toward the inner regions of the host galaxies is limited by the accuracy of our deblending technique.

\begin{figure*}
\centering
\includegraphics[width=2.0\columnwidth]{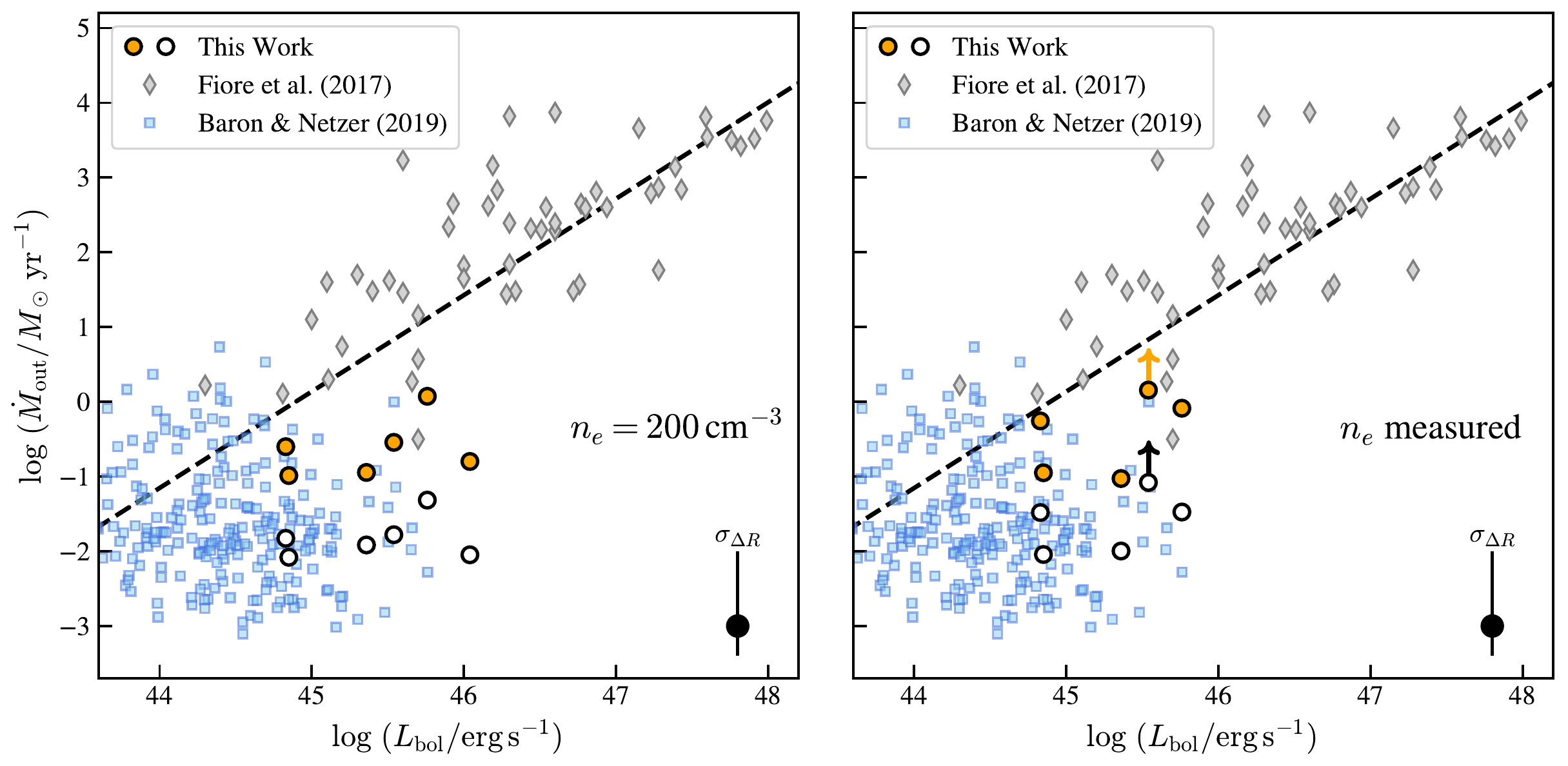}\\
\caption{\label{fig:out_Mdot} The mass outflow rate as a function of bolometric luminosity. (Left) Mass outflow rates computed by adopting $n_e=200\,$cm$^{-3}$, following \citet{Fiore2017}. The open circles show the $\dot{M}_{\rm out}$ values estimated assuming a homogeneously filled cone geometry, while the orange circles represent the $\dot{M}_{\rm out}$ values estimated by adopting the expanding shell geometry and a shell with thickness $200$\,pc. For the latter case, the vertical bar plotted in the bottom-right corner represents the systematic uncertainty associated with our assumption for the shell thickness. The dashed line shows the best fit reported by \citet{Fiore2017}. We also show the mass outflow rates given by \citet{Baron2019b} for a sample of less powerful AGNs, using their ``$\log U$'' method. (Right) Same as the left panel but using $n_e$ estimated from the observed [S\,{\sc ii}] doublet (Section~\ref{sec:outflows_obsig}). For PG\,2130+099, we only present upper limits (assuming $n_e = 40\,$cm$^{-3}$), which are highlighted by upward-pointing arrows attached to the corresponding data points. We do not show PG\,1211+143 because it has an unconstrained value of $n_e$ for the outflow. In all cases the mass outflow rates are low, $\dot{M}_{\rm out} \lesssim 1.6\, M_\odot\,$yr$^{-1}$.}
\end{figure*}

\begin{figure*}
\centering
\includegraphics[width=2.0\columnwidth]{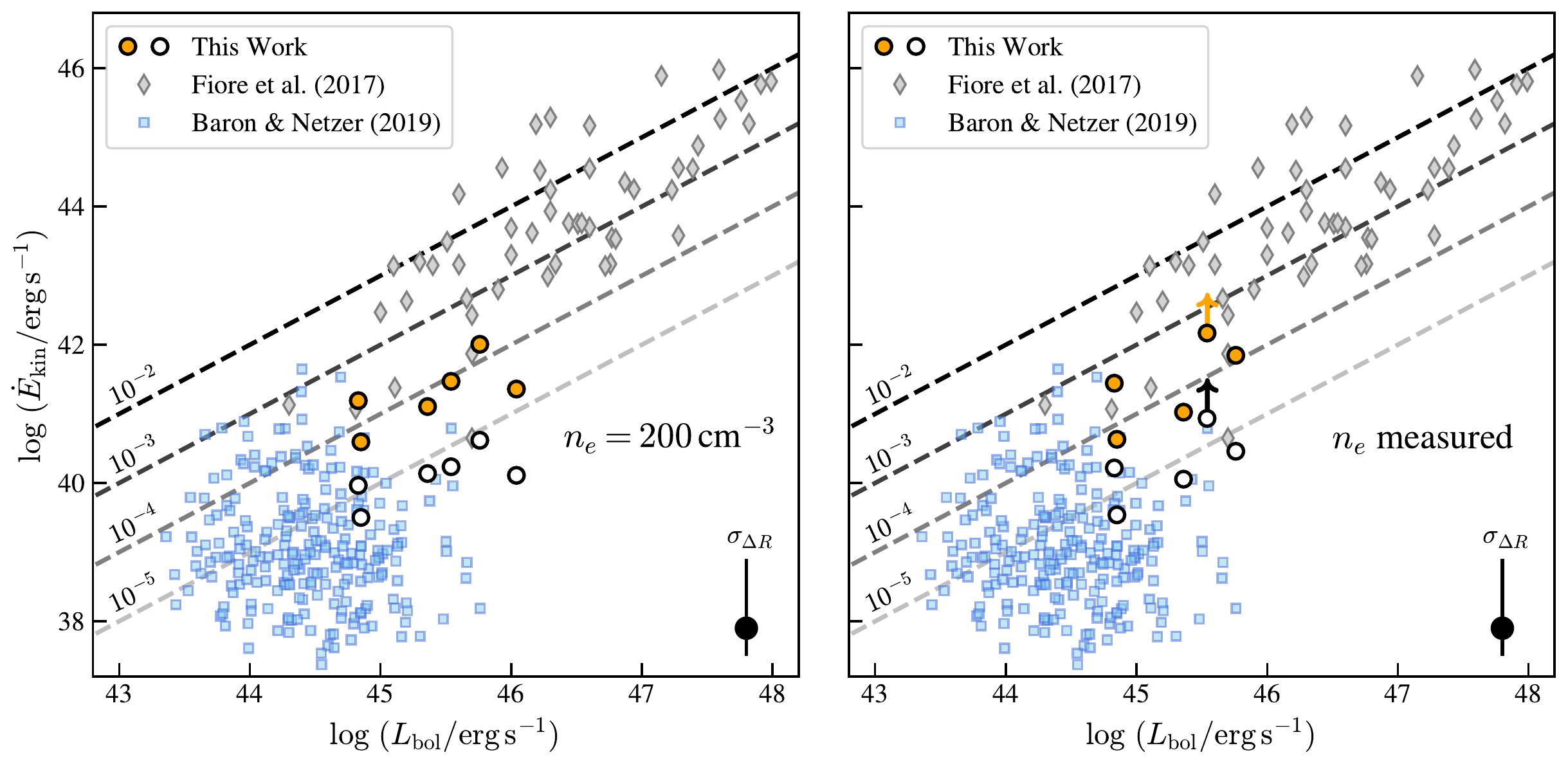}\\
\caption{\label{fig:out_Ekin} The outflow kinetic injection rate as a function of bolometric luminosity. The data are color-coded as in Figure~\ref{fig:out_Mdot}. (Left) Kinetic injection rates computed by adopting $n_e=200\,$cm$^{-3}$. The vertical bar on the bottom-right corner represents the systematic uncertainty associated with our assumed shell thickness.  The dashed lines demarcate constant ratios of $\dot{E}_{\rm kin} / L_{\rm bol}$. (Right) Same as the left panel but using values of $n_e$ estimated from the observed [S\,{\sc ii}] doublet (Section~\ref{sec:outflows_obsig}). For PG\,2130+099, we only present upper limits assuming $n_e = 40\,$cm$^{-3}$, as shown by upward-pointing arrows attached to the corresponding data points. We do not show PG\,1211+143 because it has an unconstrained value for the outflow $n_e$. Only PG\,0050+124 presents a $\dot{E}_{\rm kin} / L_{\rm bol} $ value similar to the lower bound of those reported by \citet{Fiore2017}.}
\end{figure*}

\subsubsection{Mass Outflow Rates and Energetics}
\label{sec:outflows_ME}

We now proceed to compute the mass outflow rates ($\dot{M}_{\rm out}$) and kinetic injection rates ($\dot{E}_{\rm kin}$), following \citet{Fiore2017}. Assuming that the kinematic features noted in Section~\ref{sec:outflows_obsig} correspond to genuine outflow signatures, we isolate the emission associated with those pixels in order to quantify their outflow properties (e.g., \citealt{Feruglio2020}). We select the Seyfert-classified pixels (according to the [S\,{\sc ii}]/H$\alpha$ map) along the ionization cones that have $\sigma_{V, \rm H\alpha} \geq 50$\,km\,s$^{-1}$, which lies comfortably above the velocity dispersion threshold of the star-forming regions (open circles in Figure~\ref{fig:cone_kinprofs}). For each individual pixel, we remove the parent Voronoi cell stellar continuum component (\S~\ref{sec:Stellar-comp}) from its spectrum (recall that we have already deblended the nuclear spectrum), and spatially collapse the spectra of all pixels to obtain a pure emission-line spectrum for each side of the ionization cone. To characterize the outflow spectra, we fit double-Gaussian profiles to the emission lines to account for line asymmetries. Following several authors in the literature, we compute the outflow velocity $V_{\rm out} = V_{\rm cone} + 2\, \sigma_{\rm cone}$, where $V_{\rm cone}$ and $\sigma_{\rm cone}$ are given by the best-fit H$\alpha$ line centroid and line width (e.g., \citealt{Rupke2013,Fiore2017,Feruglio2020}). The associated ionized gas mass is computed as

\begin{equation}
\label{eq:M_out}
 M_{\rm out} = 3.4 \times 10^6 \left( \frac{100\,{\rm cm^{-3}}}{n_e} \right) \left(  \frac{L_{\rm H\alpha}}{10^{41}\,{\rm erg\,s^{-1}}} \right)\,M_\odot,
\end{equation}

\noindent with $L_{\rm H\alpha}$ the extinction-corrected H$\alpha$ luminosity. Two sets of values for $M_{\rm out}$ are calculated, one using $n_e$ derived from [S\,{\sc ii}]\,$\lambda 6716$/[S\,{\sc ii}]\,$\lambda 6731$, and another assuming $n_e = 200$\,cm$^{-3}$ to make a consistent comparison with the quantities reported by \citet{Fiore2017}.\footnote{\citet{Fiore2017} scaled their ionized gas masses to this value of $n_e$ for the purposes of gathering data reported in the literature from several authors.} We derive $n_e$ values in the range of $\sim 100-360\,$cm$^{-3}$ with the exception of PG\,1211+143 and PG\,2130+099. For the former system we were unable to derive a reliable $n_e$ estimate, while for the later case we report $n_e = 40$\,cm$^{-3}$ upper limits. For a homogeneously filled cone geometry, the mass outflow rate \citep{Fiore2017} is given by,

\begin{equation}
\label{eq:Mdot_out_cone}
 \dot{M}_{\rm out} = 3 \, \left( \frac{V_{\rm out}}{100\,{\rm km\,s^{-1}}} \right)  \left( \frac{M_{\rm out}}{10^7\,M_\odot} \right) \left( \frac{1\,{\rm kpc}}{R_{\rm out}} \right)\,M_\odot\,{\rm yr^{-1}}, 
\end{equation}

\noindent where $R_{\rm out}$ is the radius at which the mass outflow rate is computed.  We set this value equal to the mean radius at which we detect the enhancement of the H$\alpha$ velocity dispersion. However, as we suggested in Section~\ref{sec:outflows_obsig}, a homogeneously filled cone geometry may not be the most suitable model to describe our data.  If an expanding shell is a better approximation, 

\begin{equation}
\label{eq:Mdot_out_shell}
 \dot{M}_{\rm out} = \left( \frac{V_{\rm out}}{100\,{\rm km\,s^{-1}}} \right)  \left( \frac{M_{\rm out}}{10^6\,M_\odot} \right) \left( \frac{100\,{\rm pc}}{\Delta R_{\rm shell}} \right)\,M_\odot\,{\rm yr^{-1}}, 
\end{equation}

\noindent where $\Delta R_{\rm shell}$ is a shell thickness that must be assumed \citep{Husemann2019a}. We use $\Delta R_{\rm shell} = 200$\,pc and adopt a range of 20--500\,pc \citep{Husemann2019a} to bracket the systematic error budget. Then, the kinetic energy injection rate is simply estimated as $\dot{E}_{\rm kin} = 0.5 V_{\rm out}^2 \dot{M}_{\rm out}$. Table~\ref{tab:derived_quantities} lists the estimates of $M_{\rm out}$, $\dot{M}_{\rm out}$, and $\dot{E}_{\rm kin}$ for the ionized gas outflows.

Figure~\ref{fig:out_Mdot} presents the mass outflow rates as a function of the AGN bolometric luminosity, $L_{\rm bol} = 10\,L_{5100}$ \citep{Richards2006}. We consider two separate cases: (1) adopting $n_e=200\,$cm$^{-3}$, as assumed by \citet{Fiore2017} and (2) using the actually measured values or upper limits of $n_e$. In both cases, we show $\dot{M}_{\rm out}$ computed assuming an outflow geometry of a homogeneously filled cone and a shell-like cone.  We find that all six of the PG quasars have very low mass outflow rates ($\dot{M}_{\rm out} \approx 0.008-0.5\,M_\odot\,$yr$^{-1}$) when compared with the sample collected by \citet{Fiore2017} at similar bolometric luminosities ($L_{\rm bol}\approx 10^{45}-10^{46}$\,erg\,s$^{-1}$), for mass outflow rates computed assuming identical $n_e$ and outflow geometry.  Our values of $\dot{M}_{\rm out}$ are similar to those reported for less powerful AGNs by \citet{Baron2019b}, who derived mass outflow rates based on the ionization parameter and AGN luminosity (their ``$\log U$'' method). We also find similar $\dot{M}_{\rm out}$ values compared to those reported by \citet{Rojas2020}, who used the nuclear [O\,{\sc iii}] line shape and luminosity to estimate $\dot{M}_{\rm out}$ for a sample of nearby X-ray-selected AGNs with bolometric luminosities comparable to those of our PG quasars. 

The mass outflow rates derived here are extraordinarily sensitive to the assumptions of the outflow geometry, which is essentially unknown. For example, the mass outflow rates increase systematically by a factor of $\sim 10$ if we adopt a shell-like outflow cone geometry, and, in view of the linear scaling with shell-thickness, another order of magnitude boost in $\dot{M}_{\rm out}$ can be achieved if $\Delta R_{\rm shell} \approx 20$\,pc. In the face of the other sources of systematic uncertainty, by lowering the H$\alpha$ velocity dispersion threshold to $\sim$30\,km\,s$^{-1}$ when selecting the Seyfert-classified pixels we find that $\dot{M}_{\rm out}$ increases only by a factor of $\sim 4$ on average, where both the increase of H$\alpha$ luminosities and the decrease of the electron densities contribute evenly to this boost factor budget. On the other hand, the two assumptions of $n_e$ that we explored do not make a big difference. However, we should recognize that density does pose a major source of uncertainty. The choice of the methodology and tracer to calculate $n_e$ affect the values of mass outflow rate and quantities that depend on it (e.g., \citealt{Baron2019b,Riffel2021b}). It is highly unlikely that a single density characterizes the line-emitting gas, nor is [S\,{\sc ii}], which has a low critical density for collisional de-excitation, the most appropriate tracer for all the gas if a large range of densities exists. \citet{Baron2019b} suggest that the [S\,{\sc ii}] emission mainly arises close behind the ionization front where most of the gas is neutral and $n_e$ is effectively lower (see also Figure~10 of \citealt{Davies2020}). The bulk of the ionized gas is poorly traced by [S\,{\sc ii}], and using this line doublet as an electron density tracer may overestimate $\dot{M}_{\rm out}$ \citep{Revalski2022}. This systematic uncertainty would also affect the mass outflow rates presented by \citet{Fiore2017} and showed in Figure~\ref{fig:out_Mdot}. Indeed, \citet{Davies2020} suggest that the  $\dot{M}_{\rm out}-L_{\rm bol}$ correlation for the more luminous AGNs weaken when correcting the data by $n_e$ (we refer to \citealt{Davies2020}, for more details).

Plotting the kinetic injection rate of the outflow versus the bolometric luminosity of the AGN (Figure~\ref{fig:out_Ekin}), we find $\dot{E}_{\rm kin}/L_{\rm bol} \lesssim 10^{-3}$, values that are systematically lower than those reported by \citet{Fiore2017} for AGNs of similar power and studied under the same model assumptions.  Our values, however, are in better agreement with the estimates reported by \citet{Baron2019b} and \citet{Rojas2020} for the less powerful AGNs and X-ray-selected AGNs with similar bolometric luminosities, respectively. Our findings suggest that ionized gas outflows with $\dot{E}_{\rm kin}/L_{\rm bol} \lesssim 10^{-3}$ are inefficient in terms of impacting the global ISM of the host galaxy.  Note that our estimates of $\dot{M}_{\rm out}$ and $\dot{E}_{\rm kin}/L_{\rm bol}$ should be regarded as upper limits if sources other than AGN-driven outflows contribute to the local enhancement in gas dispersion ($\sigma_{V, \rm H\alpha}$). The same applies regarding the use of [S\,{\sc ii}] as an electron density estimator \citep{Baron2019b,Davies2020,Riffel2021b,Revalski2022}.

\begin{table*}
	\centering
	\def\arraystretch{1.2}
	\setlength\tabcolsep{3pt}
    	\caption{\label{tab:derived_quantities} Quantities Derived in this Work}
    	\vspace{0.2mm}
	\begin{tabular}{cccccccccc}
		\hline
		\hline
		Object & $\log \, L_{[{\rm O\,III}]}$ & $R^{\rm NLR}_{\rm 16, max}$ & $R^{\rm NLR}_{\rm 16, fit}$ & $\Delta{\rm PA}_{\rm cone}$ & $\log \, M_{\rm out}$ & $\log \, \dot{M}_{\rm out}^{\rm shell}$ & $\log \, \dot{M}_{\rm out}^{\rm cone}$ & $\log \, \dot{E}_{\rm kin}^{\rm shell}$ & $\log \dot{E}_{\rm kin}^{\rm cone}$\\
		& (erg\,s$^{-1}$) & (kpc) & (kpc) & ($^\circ$) & ($M_\odot$) & ($M_\odot$\,yr$^{-1}$) & ($M_\odot$\,yr$^{-1}$) & (erg\,s$^{-1}$) & (erg\,s$^{-1}$) \\
		(1) & (2) & (3) & (4) &(5) & (6) & (7) & (8) & (9) & (10) \\
		\hline
	 PG\,0050+124 & 41.95 & $17.8\pm0.9$ & $<2.5$ & 202 & 5.5\,/\,5.7 & $-$0.1\,/\,0.1 & $-$1.5\,/\,$-$1.3 & 41.8\,/\,42.0 & 40.5\,/\,40.6 \\
         PG\,0923+129 & 41.55 & $0.7 \pm 0.5$ & $1.8 \pm 0.5$ & 206 & 5.5\,/\,5.1 & $-$0.3\,/\,$-$0.6 & $-$1.5\,/\,$-$1.8 & 41.4\,/\,41.2 & 40.2\,/\,40.0\\
         PG\,0934+013 & 41.60 & $4.7 \pm 0.5$ & $2.6 \pm 0.5$ & 182 & 4.8\,/\,4.8 & $-$0.9\,/\,$-$1.0 & $-$2.0\,/\,$-$2.1 & 40.6\,/\,40.6 & 39.5\,/\,39.5\\
         PG\,1011$-$040 & 41.36 & $2.0\pm 1.0$ & $1.6 \pm 1.0$ & \nodata & \nodata/\nodata & \nodata/\nodata & \nodata/\nodata & \nodata/\nodata & \nodata/\nodata \\
         PG\,1126$-$041 & 41.81 & $4.1 \pm 1.0$ & $4.7 \pm 1.0$ & 177 & 4.6\,/\,4.7 & $-$1.0\,/\,$-$0.9 & $-$2.0\,/\,$-$1.9 & 41.0\,/\,41.1 & 40.1\,/\,40.1 \\
         PG\,1211+143 & 41.91 & $5.6 \pm 1.6$ & $3.8 \pm 1.6$ & 210 & \nodata/\,4.7 & \nodata/\,$-$0.8 & \nodata/\,$-$2.0 & \nodata/\,41.4 & \nodata/\,40.1 \\
         PG\,1244+026 & 41.09 & $2.2 \pm 0.8$ & $2.0 \pm 0.8$ & \nodata & \nodata/\nodata & \nodata/\nodata & \nodata/\nodata & \nodata/\nodata & \nodata/\nodata \\
         PG\,1426+015 & 42.04 & $26.0 \pm 0.8$ & $4.2 \pm 0.8$ & \nodata & \nodata/\nodata & \nodata/\nodata & \nodata/\nodata & \nodata/\nodata & \nodata/\nodata \\
         PG\,2130+099 & 41.76 & $3.6\pm0.8$ & $2.9 \pm 0.8$ & 143 & $>5.7$\,/\,5.0  & $>0.2$\,/\,$-$0.5  & $>-1.1$\,/\,$-$1.8 & $>42.1$\,/\,41.5 & $>40.9$\,/\,40.2 \\
		\hline
	\end{tabular}
	\justify
	{\justify \textsc{Note}--- (1) Source name. (2) Total [O\,{\sc iii}] line luminosity. The $1\, \sigma$ error is dominated by the flux calibration uncertainty of $\sim$3--5\% \citep{Weilbacher2020}. (3) Narrow-line region maximum extension at a surface brightness value of 10$^{-16}$\,erg\,s$^{-1}$\,cm$^{-2}$\,arcsec$^{-2}$\,(1+$z$)$^{-4}$. We adopt the PSF radial size (FWHM/2) as the $1\,\sigma$ uncertainty. (4) Narrow-line region best-fit radius at the surface brightness value of 10$^{-16}$\,erg\,s$^{-1}$\,cm$^{-2}$\,arcsec$^{-2}$\,(1+$z$)$^{-4}$. We assume the PSF radial size (FWHM/2) as the $1\,\sigma$ error estimate. (5) Difference between the ionization cone sides position angles. The $1\, \sigma$ uncertainty is 16$^\circ$. (6) Outflow mass. (7) Mass outflow rate computed assuming an expanding shell-like shock front geometry. (8) Mass outflow rate estimated assuming a homogeneously filled cone geometry. (9) Outflow kinetic injection rate computed assuming expanding shell-like shock front geometry. (10) Outflow kinetic injection rate estimated assuming homogeneously filled cone geometry. As a second option, we also provide the outflow properties assuming $n_e = 200\,$cm$^{-3}$ (e.g., \citealt{Fiore2017}). Formal error estimates are not given, as systematic uncertainties dominate the error budget for these quantities.}
\end{table*}

\section{Summary and Discussion}
\label{sec:Discussion}

The role of AGN feedback on the evolution of galaxies remains a topic of intense debate. Theoretical arguments abound concerning the need for negative AGN feedback to solve a suite of otherwise intractable problems in galaxy evolution. Theoretical and numerical works argue that outflows with kinetic power of a few percent the energy radiated by the AGN ($\dot{E}_{\rm kin} / L_{\rm bol} \approx 0.05-0.5$) can significantly impact the evolution of the host galaxy \citep{DiMatteo2005,Springel2005,HopkinsQuataert2010,ZubovasKing2012,Hopkins2016}. Indeed, literature estimates based on observations of ionized, molecular, and X-ray tracers collected by \citet{Fiore2017} suggest that the kinetic power of outflows lie mainly in the range $\dot{E}_{\rm kin} / L_{\rm bol} \lesssim 0.1$, with only a few reaching values as high as $\dot{E}_{\rm kin}/L_{\rm bol} \approx 0.1-1$ (see also \citealt{Shimizu2019,Kakkad2022}). A major source of uncertainty pertains to the degree to which outflows actually couple to the cold gas reservoir of the host galaxy. If the outflow expands in a direction perpendicular to the disk plane of the galaxy, then the cold gas, which resides largely in the disk, would be hardly touched \citep{Gabor2014}. But the accretion disk is not typically aligned with the host galaxy \citep{Wilson1994}.

While there has been much tantalizing evidence of multi-phase outflows in AGNs reported in the literature (e.g., \citealt{Cicone2014,Feruglio2015,Karouzos2016,Morganti2017,Cicone2018,Fluetsch2019,Veilleux2020}), we have a poor understanding of how efficiently the outflows actually affect the host galaxy (e.g., \citealt{Harrison2018,Kakkad2022}). The naive expectation that AGN feedback expels and depletes the cold gas reservoir is certainly not borne out by much of the evidence. A variety of observations have demonstrated that active galaxies, covering a wide gamut of accretion power and evolutionary states, are categorically not gas-deficient compared to inactive, star-forming of similar stellar mass (e.g., \citealt{Bertram2007,Krips2012,Xia2012,Husemann2017,Shangguan2018,Shangguan2020b,Shangguan2019,Jarvis2020,Yesuf2020,Koss2021,Salvestrini2022}). The cold gas shows no signs of being substantially disturbed in terms of its global kinematics (e.g., \citealt{Ho2008,Shangguan2020b,Molina2021}), nor does it have any trouble forming stars (e.g., \citealt{Bernhard2019,Grimmett2020,Jarvis2020,Kirkpatrick2020,Xie2021,Zhuang2021}).

This study presents MUSE IFU observations of nine nearby ($z \lesssim 0.1$) Palomar-Green quasars that map the host galaxies at kpc-scale resolution. We designed a procedure to deblend the bright AGN from the underlying emission of the host galaxy, giving us a pixel-wise characterization of the emission lines. We use optical line-intensity ratio diagnostic diagrams to determine the underlying excitation mechanism of the line emission, and we measure the kinematics of the emission lines to estimate the physical properties of the ionized gas outflows. Our main results are:

\begin{itemize}
\item The NLR is spatially resolved for all the host galaxies. Down to an [O\,{\sc iii}] surface brightness level of $10^{-16}$\,erg\,s$^{-1}$\,cm$^{-2}$\,arcsec$^{-2}$, the NLR has radial extensions in the range $1.6-4.7$\,kpc, which correlate with the total extinction-corrected [O\,{\sc iii}] luminosity. In some cases, the [O\,{\sc iii}] emission can reach distances beyond the physical extent of the host galaxy. 

\item The radiation field of the AGN ionizes the host galaxy ISM in preferred directions, producing an NLR morphology that resembles a bipolar ionization cone.

\item The velocity dispersion of the ionized gas, as traced by H$\alpha$, can reach values $\gtrsim 100$\,km\,s$^{-1}$ at specific galactocentric radii (mostly $\sim 5-10$\,kpc) across the AGN ionization cones.  These velocity dispersions are significantly higher than the average value ($\lesssim 40$\,km\,s$^{-1}$) measured from star-forming regions located at similar distances from the galaxy center.

\item Assuming that the local velocity dispersion enhancements reflect the interaction between the AGN-driven outflow and the host galaxy ISM, we compute ionized gas mass outflow rates $\dot{M}_{\rm out} \approx 0.008-1.6\,M_\odot\,$yr$^{-1}$ and kinetic injection powers $\dot{E}_{\rm kin}/L_{\rm bol} \lesssim 10^{-3}$. These values are substantially lower than those expected from the empirical correlations of \citet{Fiore2017}, but they agree better with the estimates of \citet{Baron2019b} and \citet{Rojas2020} for AGNs with similar and lower bolometric luminosities.
\end{itemize}

Our IFU analysis of the present sample of PG quasars offers some clues on the conundrum of why the host galaxies of luminous AGNs in the local Universe seem to show so little evidence of negative AGN feedback.  Although the NLRs follow the expected luminosity-size relation, implying that the AGN radiation can permeate across a solid angle not significantly misaligned with respect to the disk plane of the host, the emission detected across the host galaxy is faint and seems to be easily swamped by the H$\alpha$ emission generated by the star formation activity (Figure~\ref{fig:polar_maps}). More importantly, the extremely low kinetic powers of $\dot{E}_{\rm kin}/L_{\rm bol} \lesssim 10^{-3}$ suggest that the ionized gas outflows are in general very poorly coupled to the ISM of the host galaxy. This casts serious doubt on the efficiency of negative AGN feedback and its global impact on the host galaxies of nearby quasars. Although the present results only pertain to the ionized gas in a small number of objects, we note that \cite{Shangguan2020b} also placed strong limits on the absence of cold, molecular outflows for a subset of the low-redshift PG quasars. \citet{Koss2021} found similar results when analyzing the molecular gas content in a large sample ($\sim 200$) of nearby X-ray-selected AGNs. 

Aside from the low effectiveness of negative AGN feedback, an intriguing characteristic revealed by the IFU observations is the prevalence of enhanced velocity dispersion for the ionized gas at certain zones located within the ionization cone. The increase in velocity dispersion is associated, at least in one instance (PG\,0934+013; Figure~\ref{fig:cone_kinprofs}), with an accompanying increase in electron density. A local increase in the velocity widths of the nebular emission lines has been observed also in the inner kpc-scale regions of nearby AGNs \citep{Ruschel2021}, whose kinematic disturbance has been regarded as an indicator of outflows within the NLR \citep{Kang2017,SunAGN2018,Husemann2019a}. Jet-like radio emission has also been reported \citep{Husemann2013,Villar2017,Santoro2020,Venturi2021}. The available radio data ($\sim 0.685-5$\,GHz) for the PG quasars suggest unresolved ($\lesssim 0\farcs5$) low-powered ratio jet emission in PG\,1011$-$040, PG0934+013 and PG1426+015, while non-conclusive reports can be made for the other host galaxies as the radio fluxes could also be produced by the optically thin synchrotron emission from AGN- and/or starburst-driven winds \citep{Silpa2020}. We interpret the localized velocity dispersion enhancements as a signature of the shock front interface between the ISM and an expanding, shell-like outflow propagating in a direction offset from the disk midplane of the host galaxy. 

\acknowledgments{
We thank to the anonymous referee for her/his helpful comments and suggestions. We acknowledge support from the National Science Foundation of China (11721303, 11991052, 12192220, 12192222, 12011540375), the China Manned Space Project (CMS-CSST-2021-A04, CMS-CSST-2021-A06) and the National Key R\&D Program of China (2016YFA0400702). We acknowledge support from FONDECYT Regular 1190818 ( F.\,E.\,B., E.\,T.) and 1200495 (F.\,E.\,B., E.,T.); ANID grants CATA-Basal AFB-170002 (F.\,E.\,B., E.\,T.), ACE210002 (F.\,E.\,B., E.\,T.) and FB210003 (F.\,E.\,B., E.\,T.); Millennium Nucleus NCN19\_058 (E.\,T.); and Millennium Science Initiative Program--ICN12\_009 (F.\,E.\,B.). This research has made use of the services of the ESO Science Archive Facility, and based on observations collected at the European Organization for Astronomical Research in the Southern Hemisphere under ESO programme IDs 094.B$-$0345(A), 095.B$-$0015(A), 097.B$-$0080(A), 0101.B$-$0368(B), 0103.B$-$0496(B) and 0104.B$-$0151(A).}

\software{\textsc{Astropy}\,\citep{astropy:2013,astropy:2018}, \textsc{lmfit}\,\citep{Newville2014}, \textsc{matplotlib}\,\citep{Hunter2007}, \textsc{numpy}\,\citep{Oliphant2006}, \textsc{photutils}\,\citep{Bradley2020}, \textsc{scikit-image}\,\citep{scikit-image}, \textsc{scipy}\,\citep{scipy2020}.}

\appendix

\bigskip
\bigskip
\bigskip
\bigskip
\bigskip
\bigskip
\bigskip
\bigskip
\bigskip
\bigskip
\bigskip

\section{Two-dimensional Maps for Individual Host Galaxies}
\label{sec:AppA}

Figure~\ref{fig:all_maps} provides the set of two-dimensional maps for the full sample.

\begin{figure*}
\figurenum{A1}
\centering
\includegraphics[width=2.0\columnwidth]{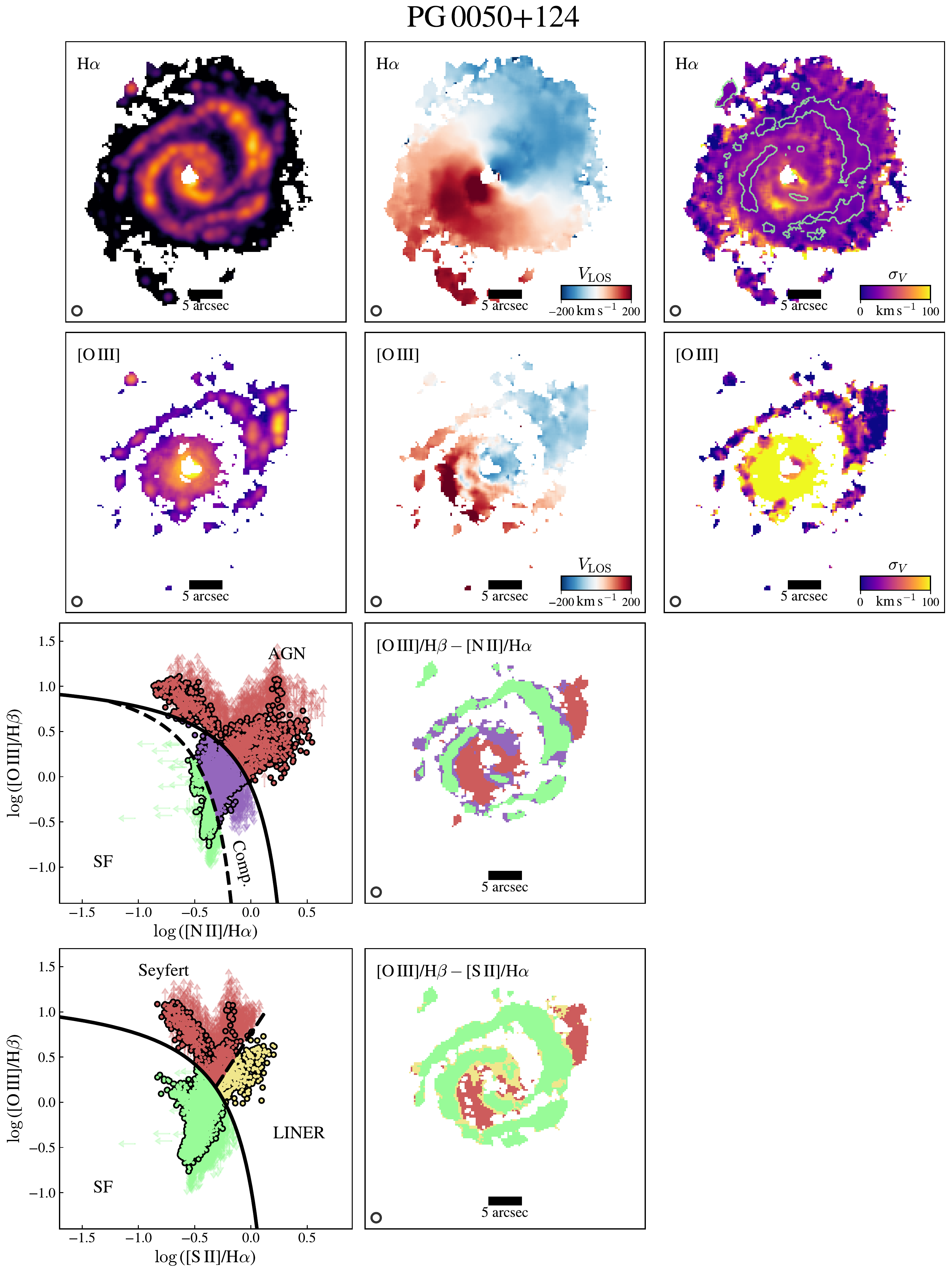}\\
\caption{\label{fig:all_maps} Two-dimensional moment maps and ionization diagnostic diagrams for the PG quasar host galaxies.  The panels are organized and color-coded following Figure~\ref{fig:example_maps}.}
\end{figure*}

\begin{figure*}
\centering
\figurenum{A1}
\includegraphics[width=2.0\columnwidth]{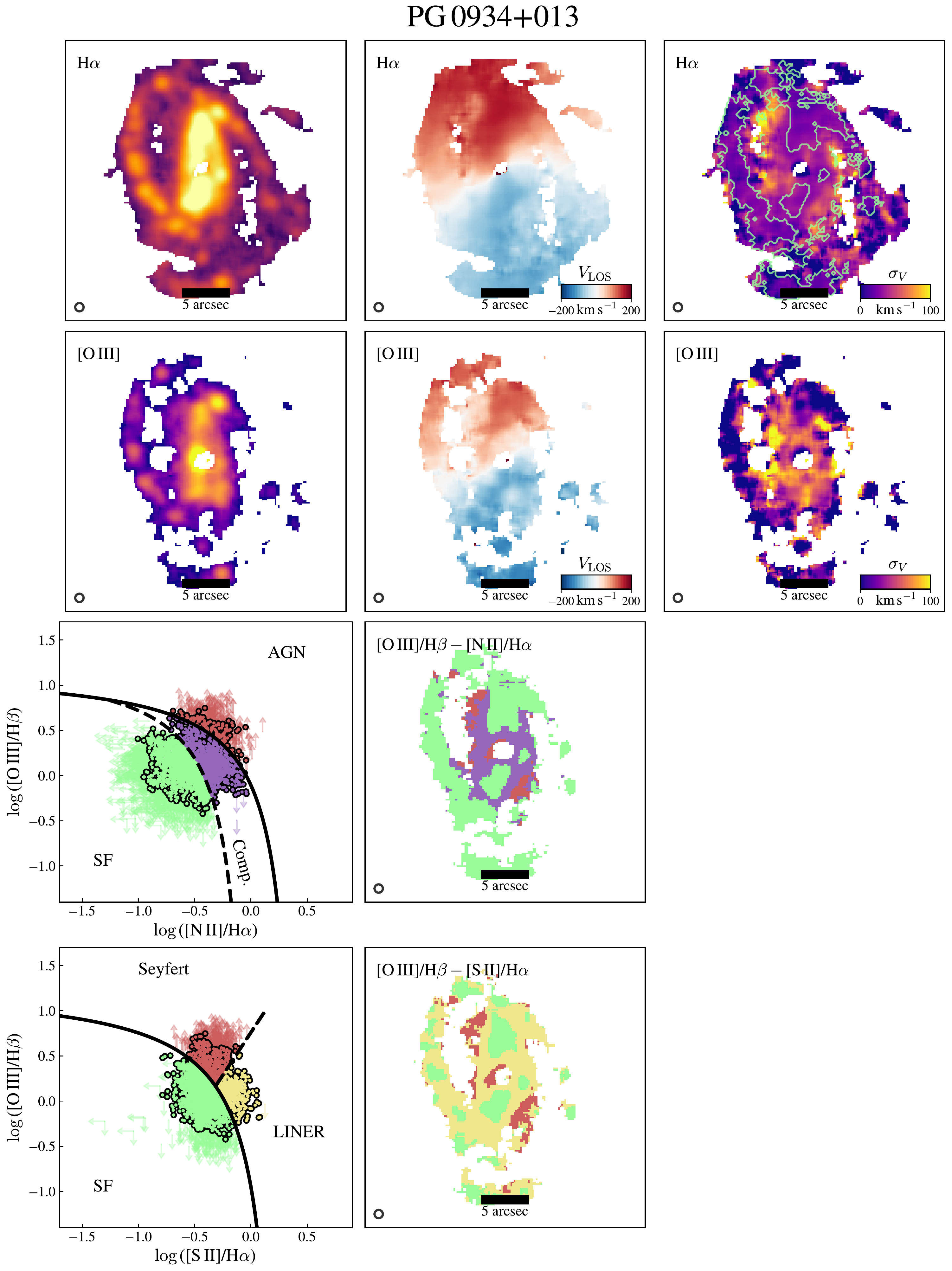}\\
\caption{Continued.}
\end{figure*}

\begin{figure*}
\centering
\figurenum{A1}
\includegraphics[width=2.0\columnwidth]{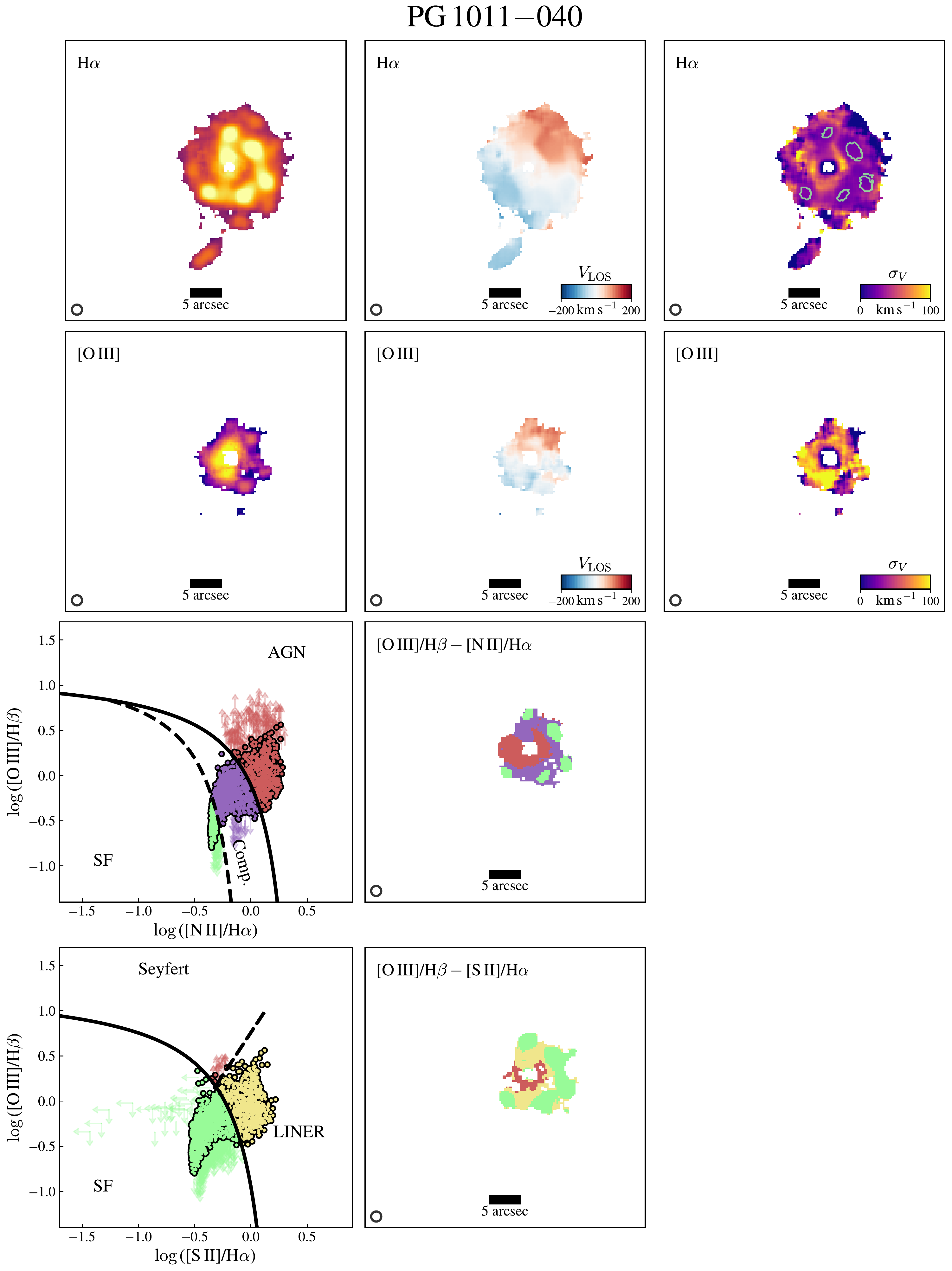}\\
\caption{Continued.}
\end{figure*}

\begin{figure*}
\centering
\figurenum{A1}
\includegraphics[width=2.0\columnwidth]{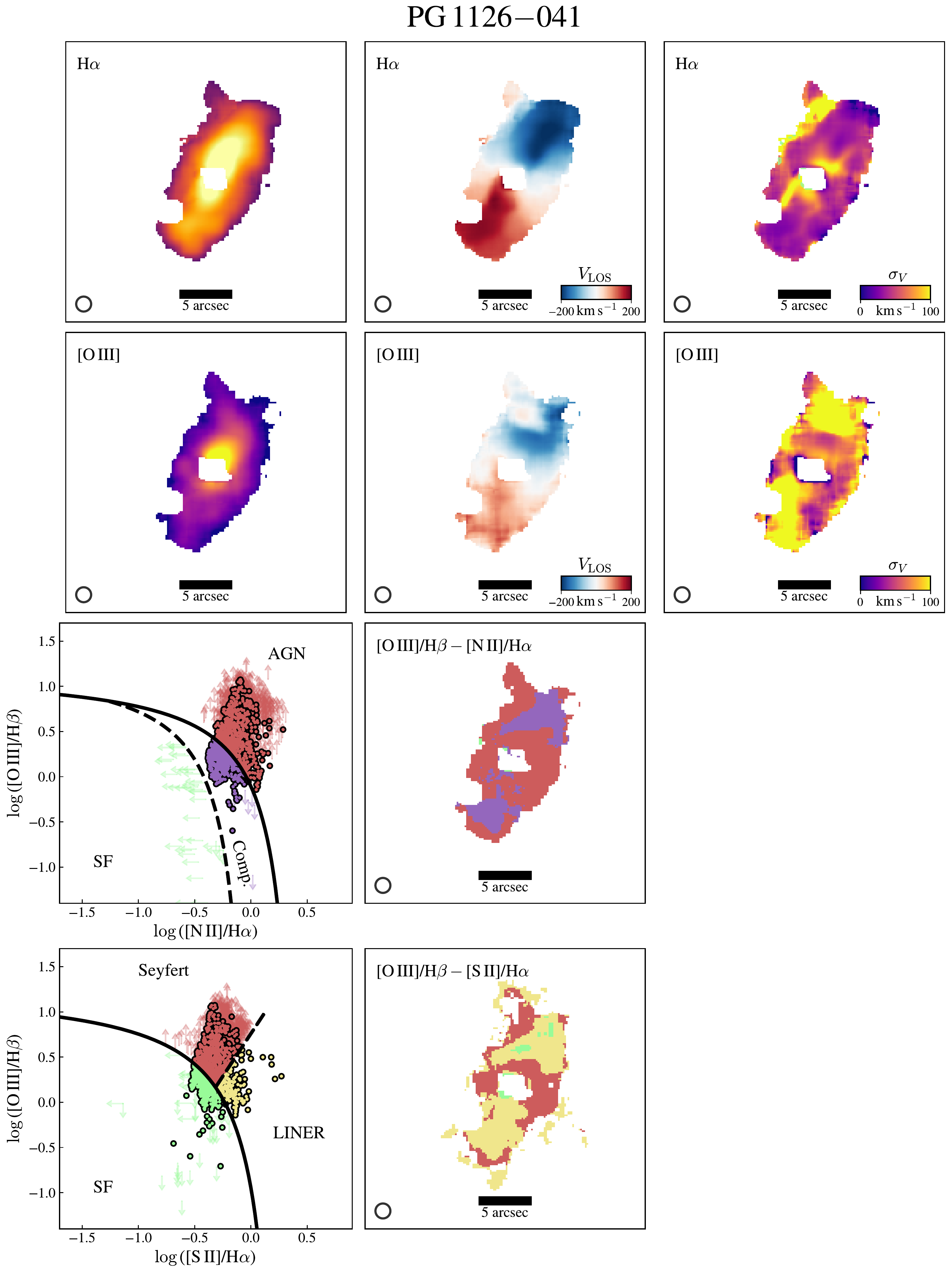}\\
\caption{Continued.}
\end{figure*}

\begin{figure*}
\centering
\figurenum{A1}
\includegraphics[width=2.0\columnwidth]{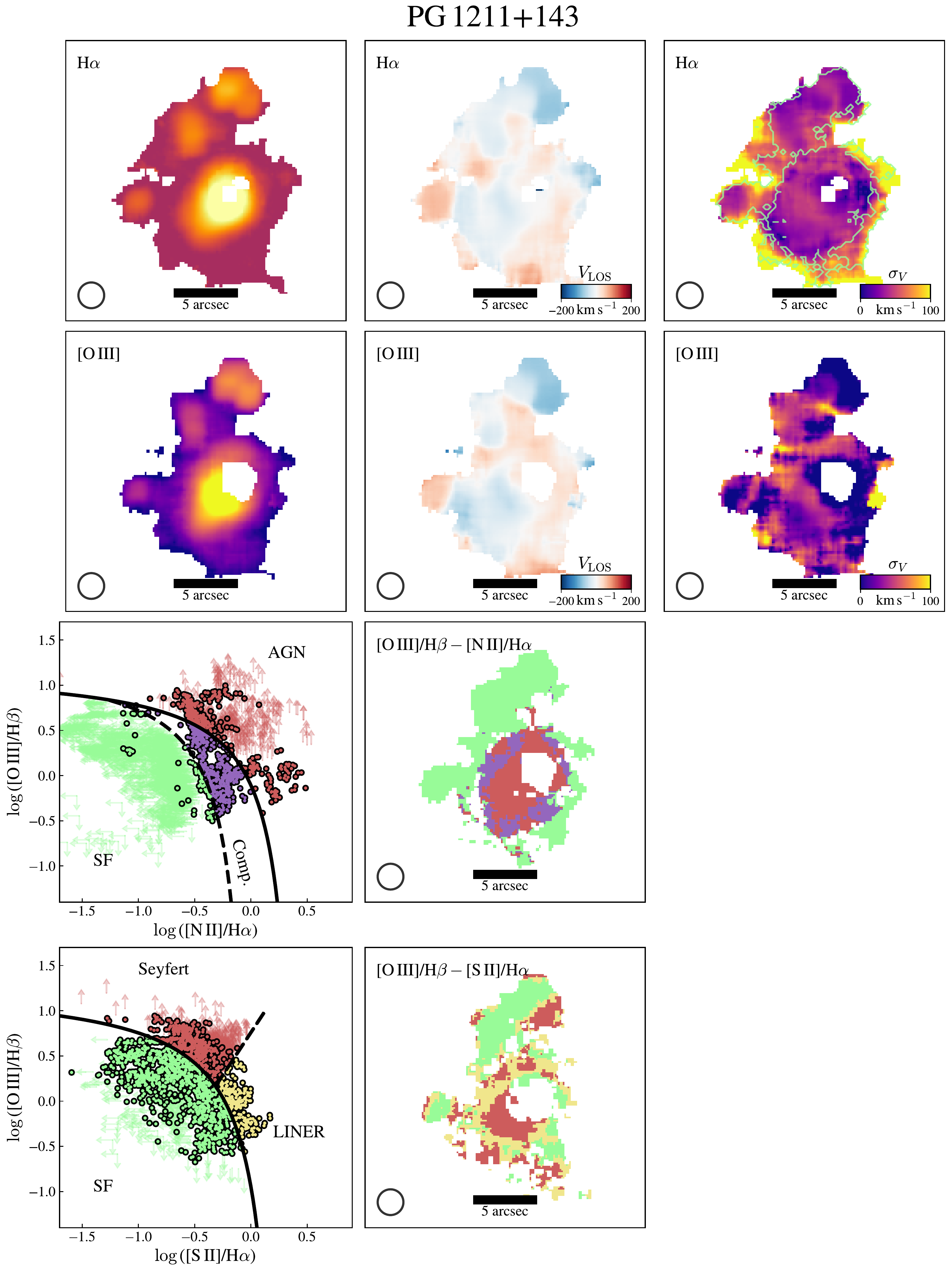}\\
\caption{Continued.}
\end{figure*}

\begin{figure*}
\centering
\figurenum{A1}
\includegraphics[width=2.0\columnwidth]{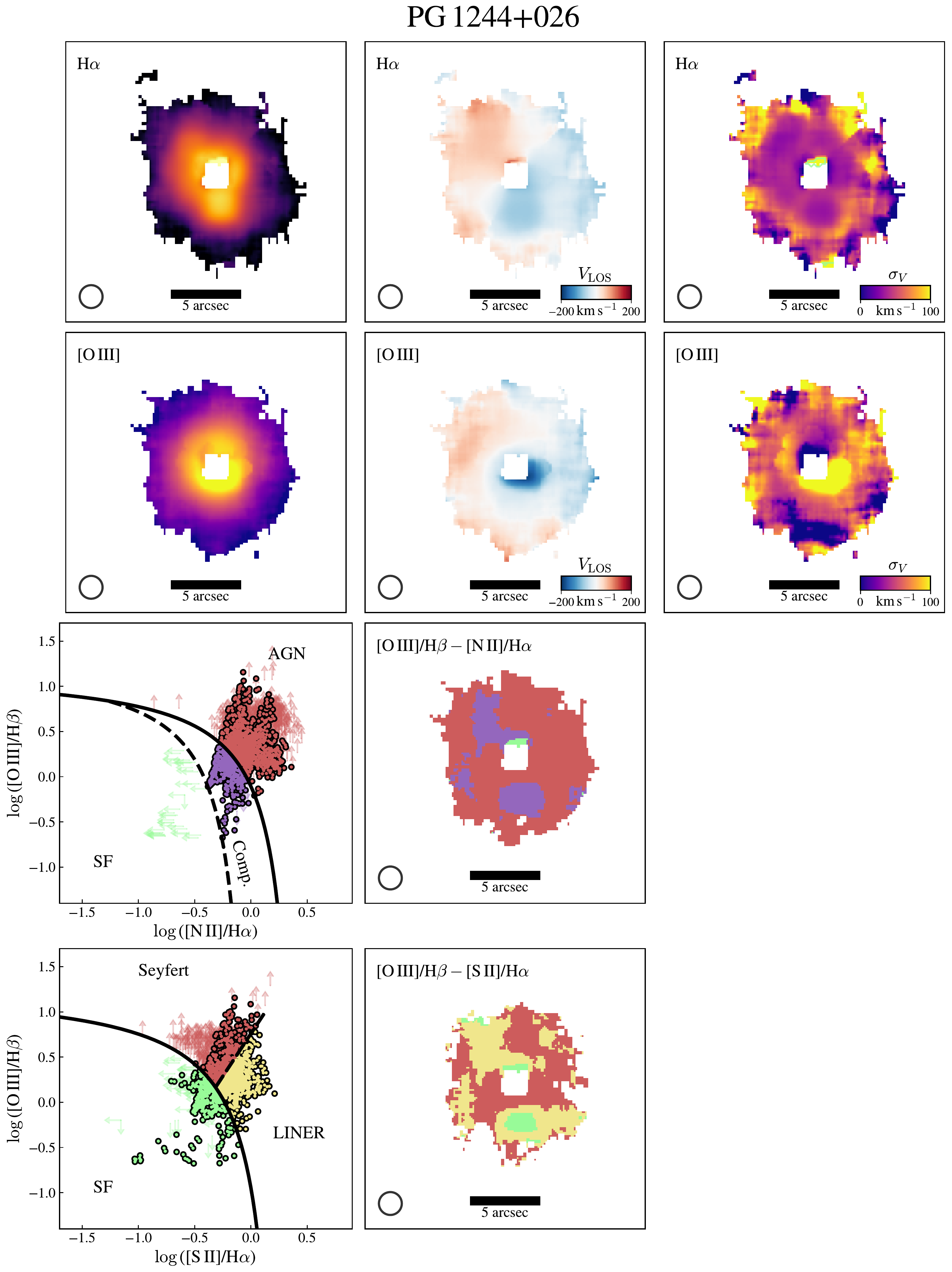}\\
\caption{Continued.}
\end{figure*}

\begin{figure*}
\centering
\figurenum{A1}
\includegraphics[width=2.0\columnwidth]{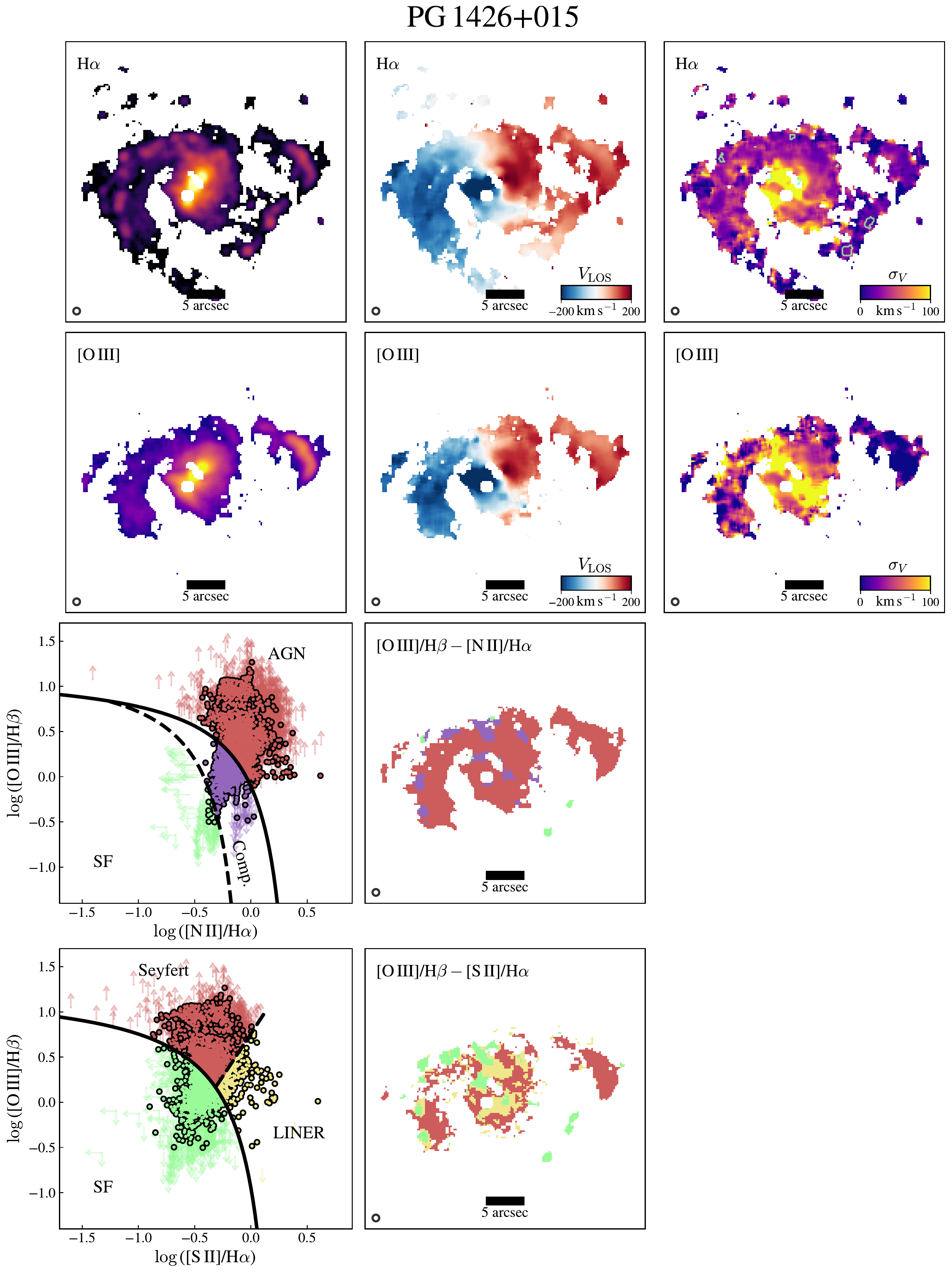}\\
\caption{Continued.}
\end{figure*}

\begin{figure*}
\centering
\figurenum{A1}
\includegraphics[width=2.0\columnwidth]{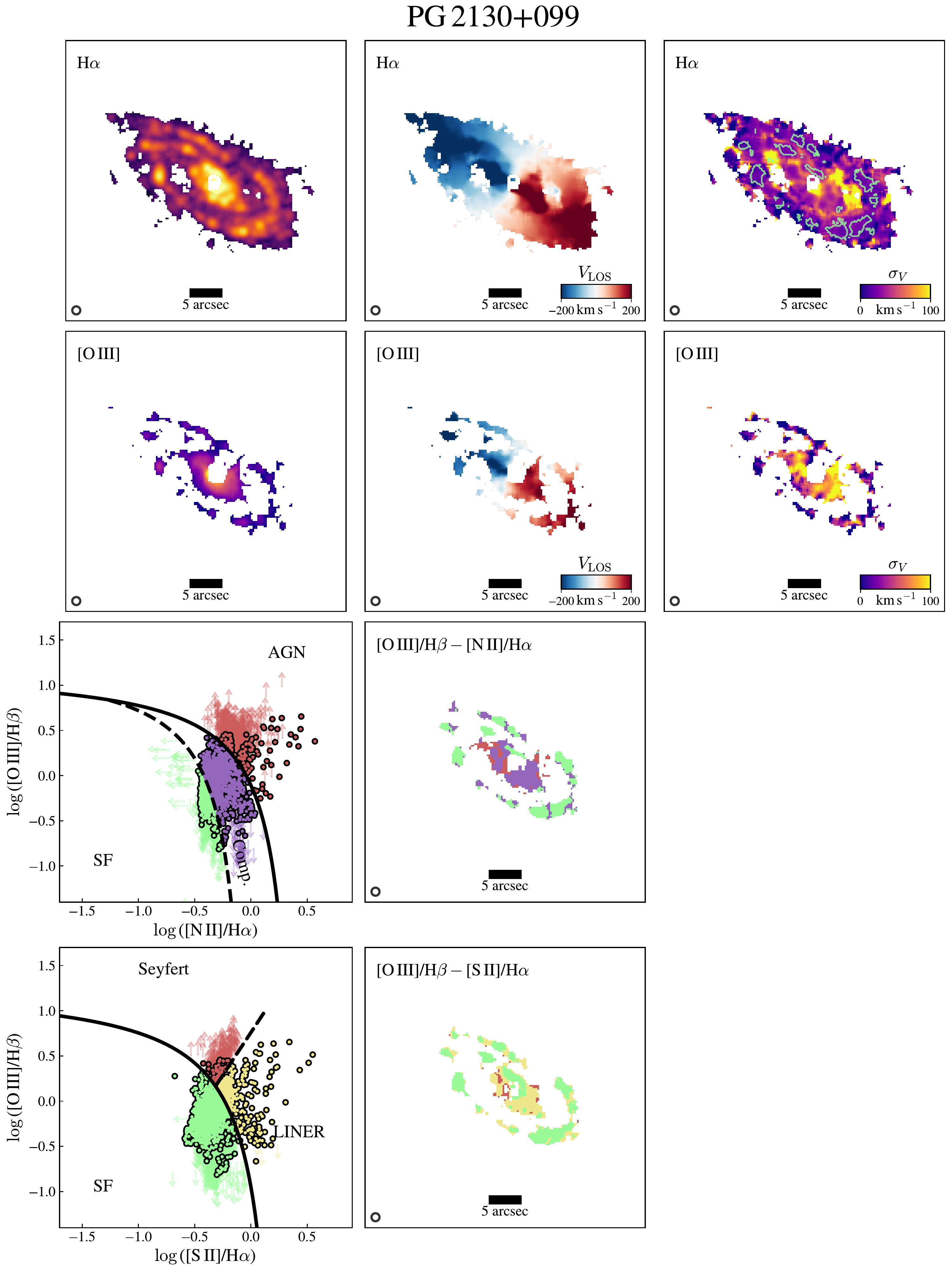}\\
\caption{Continued.}
\end{figure*}

\vfill\eject

\section{Polar Maps and Covering Fraction Azimuthal Profiles for Individual Host Galaxies}
\label{sec:AppB}

Figure~\ref{fig:polar_maps_app} provides the set of polar maps and covering fraction azimuthal profiles for the full sample.

\begin{figure*}
\figurenum{B1}
\centering
\includegraphics[width=1.8\columnwidth]{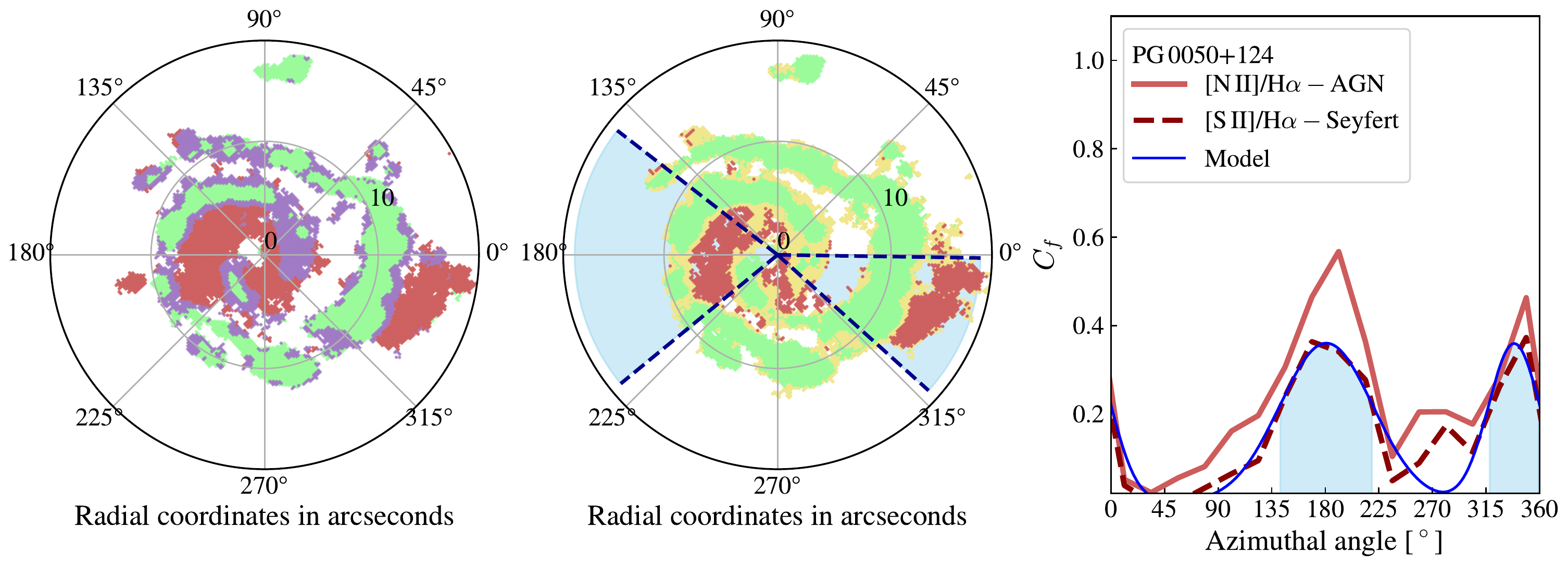}\\
\includegraphics[width=1.8\columnwidth]{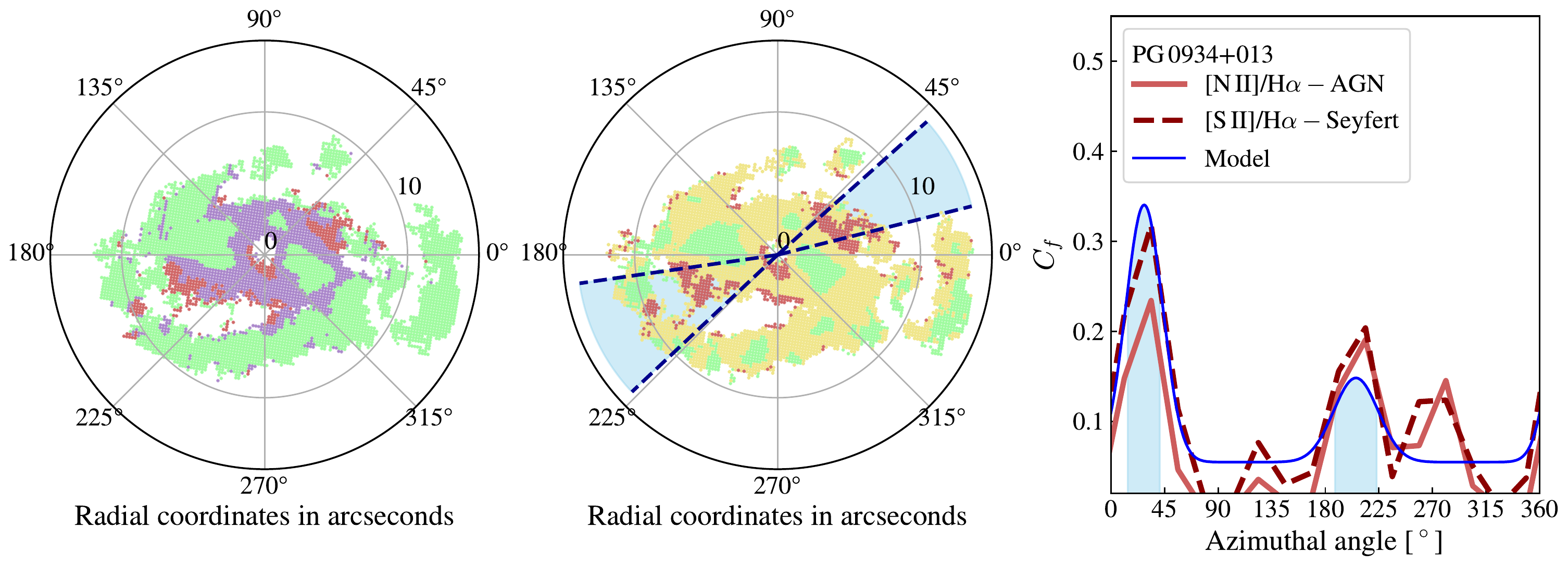}\\
\includegraphics[width=1.8\columnwidth]{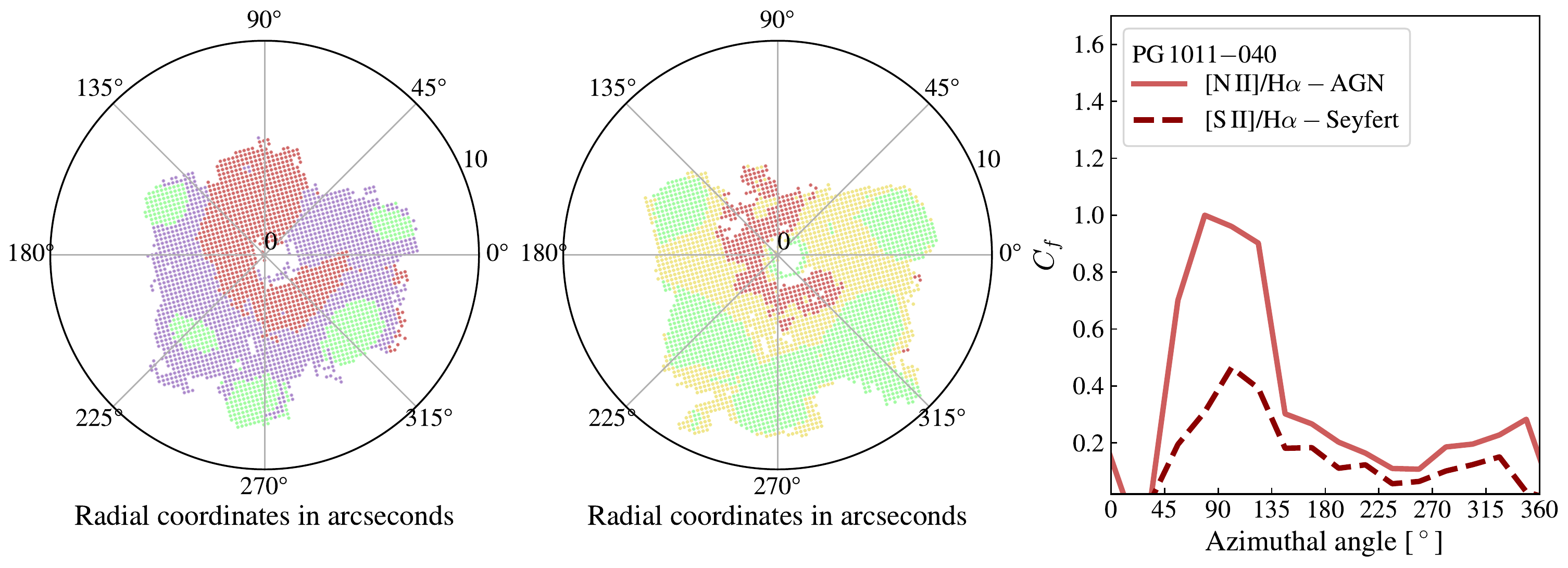}\\
\includegraphics[width=1.8\columnwidth]{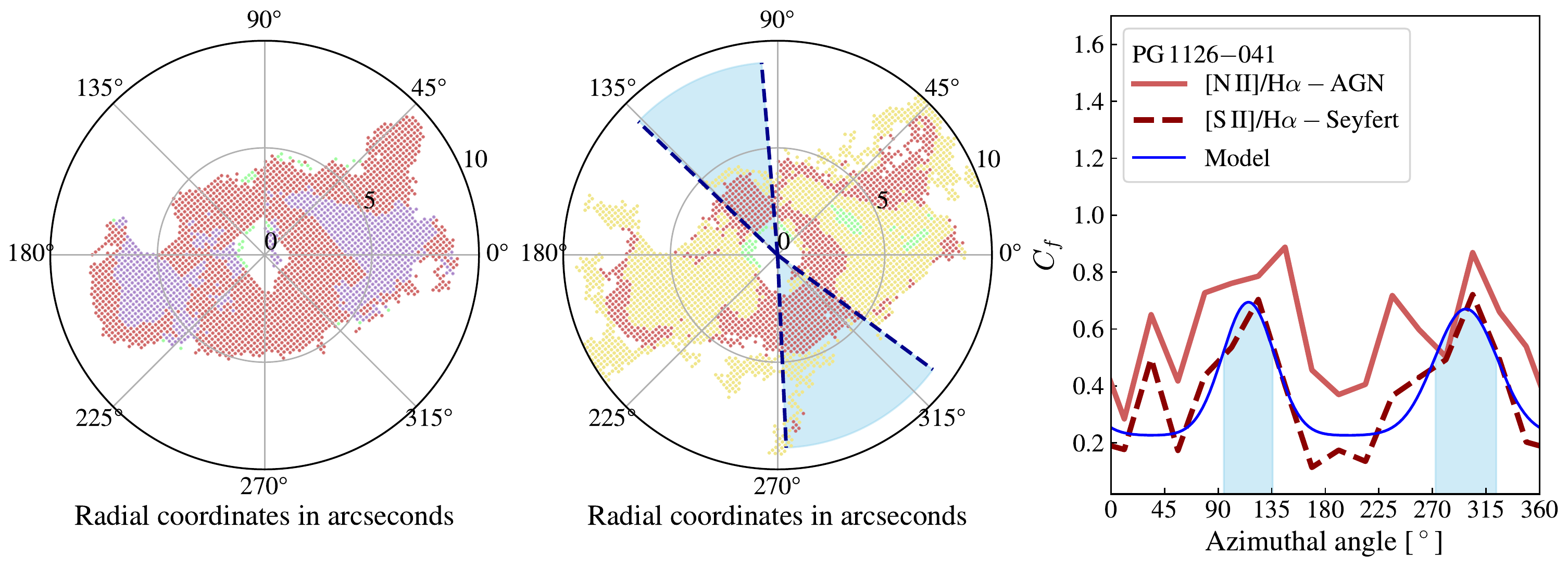}\\
\caption{\label{fig:polar_maps_app} The host galaxies of PG quasars seen in polar projection. The panels are organized and color-coded following Figure~\ref{fig:polar_maps}.}
\end{figure*}

\begin{figure*}
\figurenum{B1}
\centering
\includegraphics[width=1.8\columnwidth]{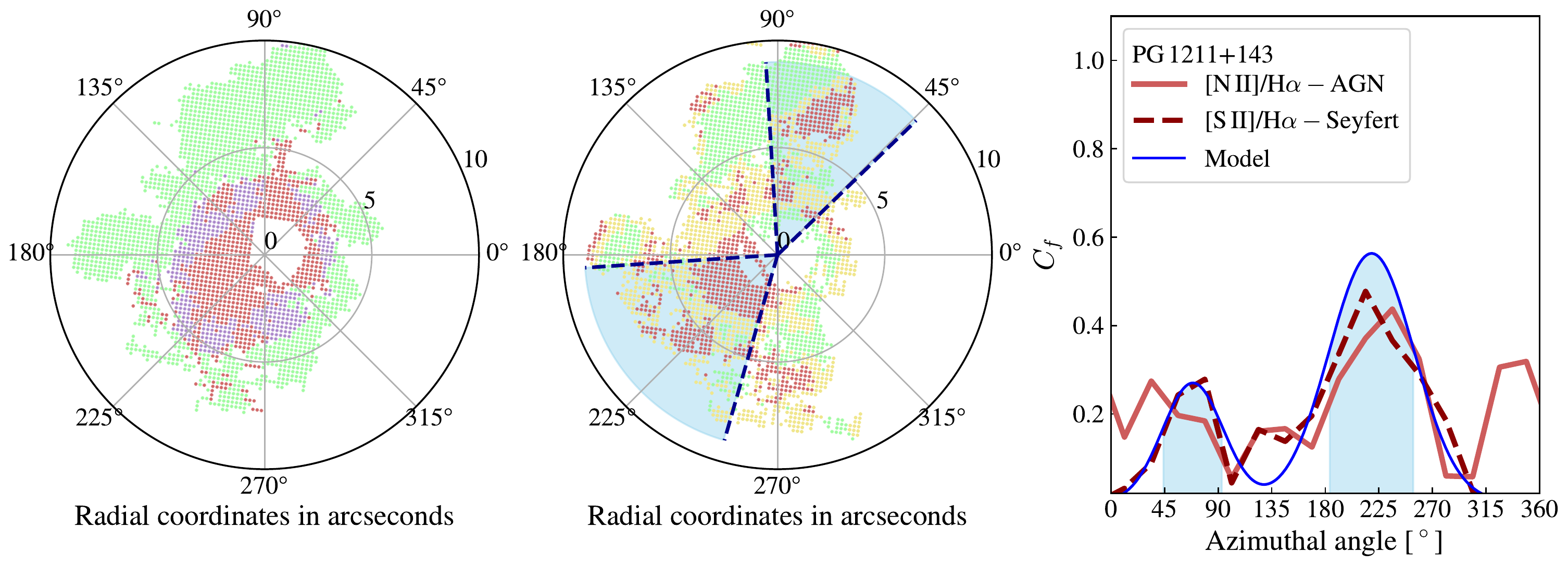}\\
\includegraphics[width=1.8\columnwidth]{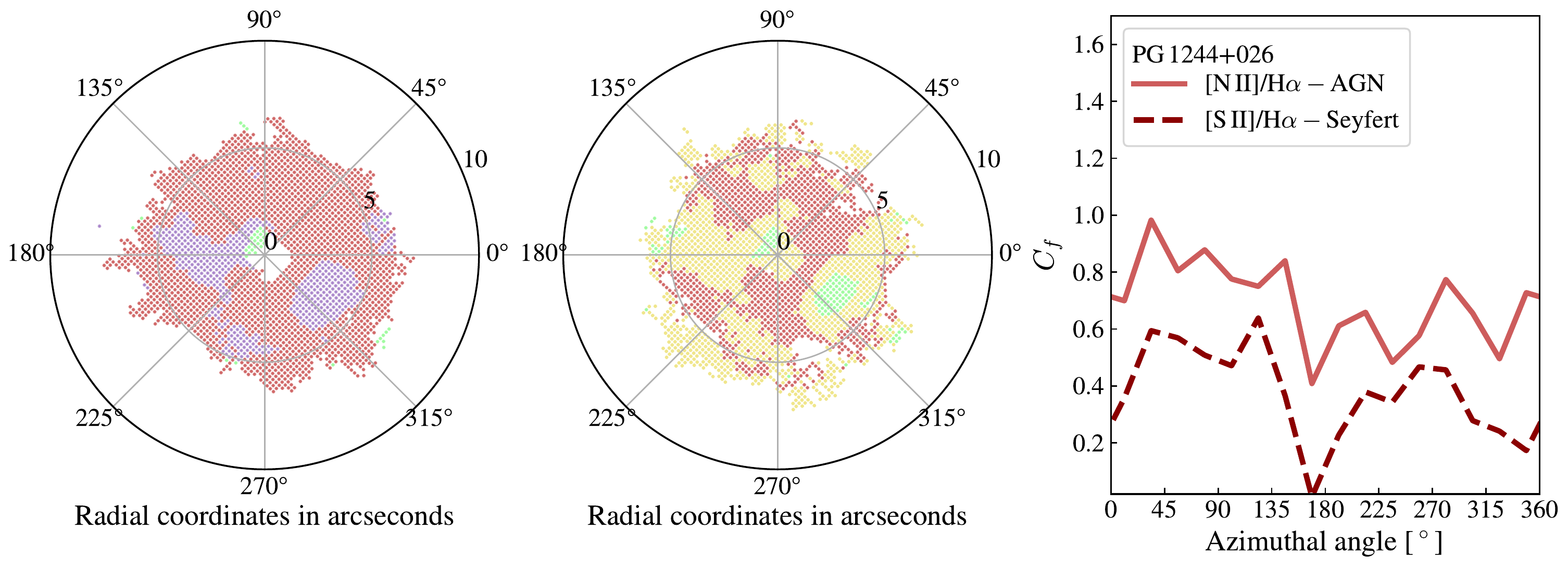}\\
\includegraphics[width=1.8\columnwidth]{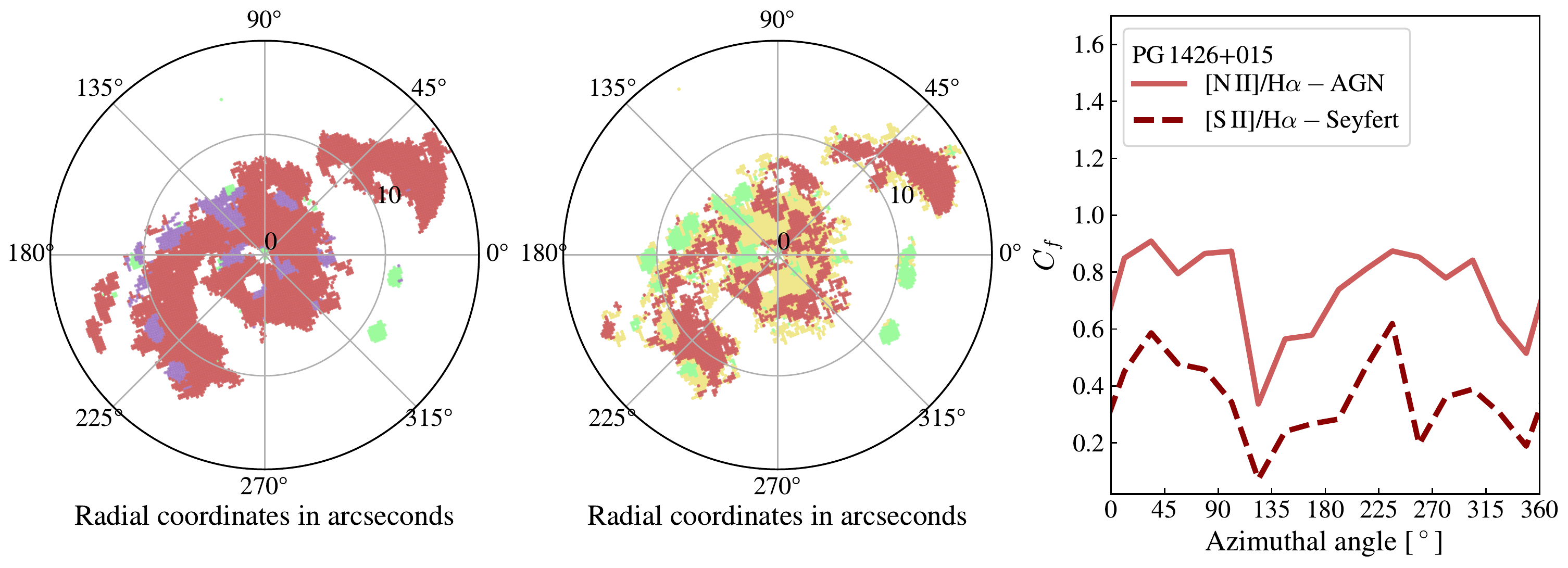}\\
\includegraphics[width=1.8\columnwidth]{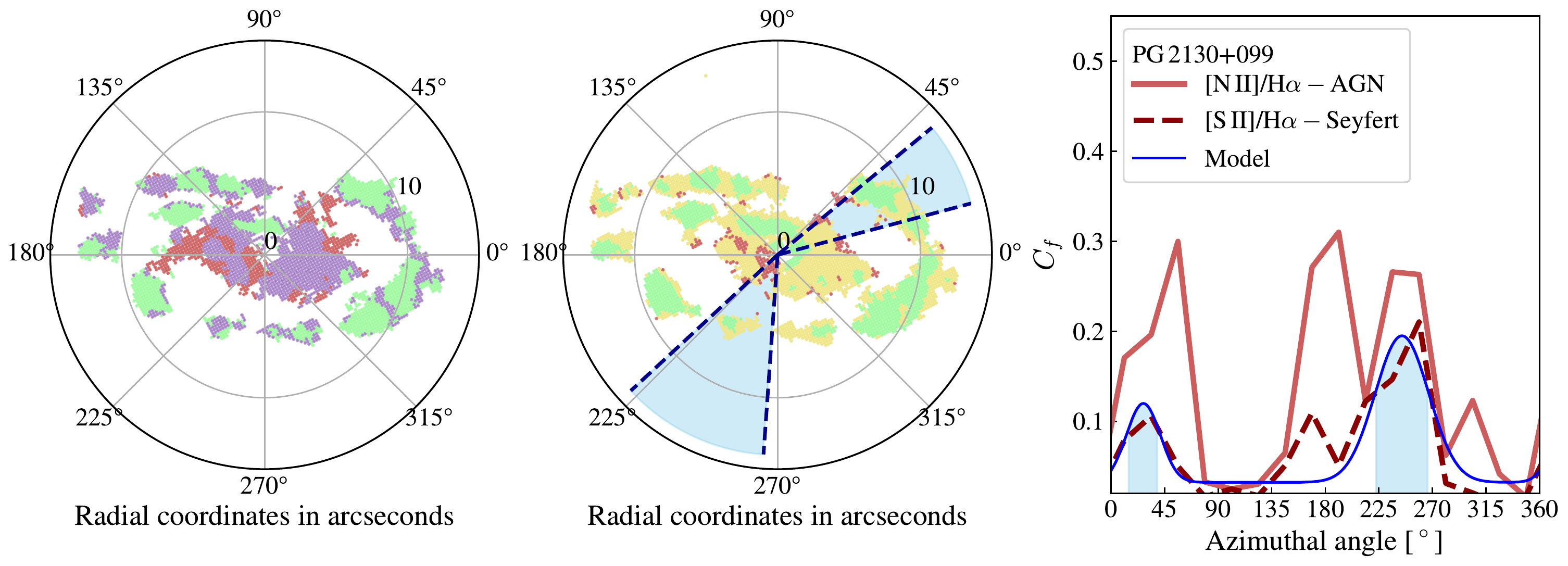}\\
\caption{Continued.}
\end{figure*}

\vfill\eject
\section{Notes for Individual Host Galaxies}
\label{sec:AppC}

\begin{itemize}
\item[]\textbf{PG\,0050+124}: The host galaxy has two extended and clumpy spiral arms that extend out to the outskirts of the stellar disk. The emission coming from the spirals can be unambiguously associated with underlying star formation activity according to the BPT maps. On the other hand, the emission coming from the inter-arm regions are much fainter and likely produced by the AGN ionization and/or shocks. The LOS velocity map clearly shows a disk-like rotational pattern, while the velocity dispersion map suggests more complex ISM dynamics. The spiral arms can be associated with regions that show lower values of $\sigma_{V,{\rm H\alpha}}$ within the host galaxy, while the inter-arm regions tend to show larger $\sigma_{V,{\rm H\alpha}}$. 

\item[]\textbf{PG\,0923+129}: The H$\alpha$ emission exhibits an inner disk structure from which two clumpy spiral arms extend out to the outskirts of the host galaxy, which shows ubiquitous star formation activity. The inner H$\alpha$ emission seems to trace a ring-like structure. This is interesting, as the galaxy appears to be an S0 system when seen in HST or MUSE broad-band images. A main region with Seyfert classification seen toward the south-east suggests the presence of an ionization cone. The velocity map shows the characteristic rotation pattern of a disk galaxy, while the velocity dispersion map suggests that star formation occurs in zones with lower $\sigma_{V,{\rm H\alpha}}$. 

\item[]\textbf{PG\,0934+013}: Two outer, clumpy, spiral arm-like regions can be clearly seen in the H$\alpha$ map. The central zone, where a bar is present, also emits H$\alpha$, which, according to the BPT maps is powered by star formation activity. The zones ionized by the AGN are mainly distributed from the north-east to the south-west in the form of a biconical ionization cone. The [S\,{\sc ii}]/H$\alpha$ BPT diagram, however, suggests that LINER-like emission is ubiquitous across the host galaxy. Similar to PG\,0050+124 and PG\,0923+129, regions with LINER characteristics have higher $\sigma_{V,{\rm H\alpha}}$ than those dominated by star formation. The LOS velocity map shows a clear disk-like rotation pattern. 

\item[]\textbf{PG\,1011$-$040:} The H$\alpha$ emission traces two main zones, a nuclear region surrounded by ring-like clumpy structure. Five large star-forming clumps are clearly seen inside the ring. The LOS velocity field is complex, but the velocity gradient varies smoothly across the field-of-view, suggesting that both structures are spatially connected. This is also supported by the rest-frame $I$-band HST/WF3 image and MUSE white-light image, which reveals a stellar bar crossing from the ring-like structure from one extreme to the other through the nuclear region. The central zone has larger $\sigma_{V,{\rm H\alpha}}$.

\item[]\textbf{PG\,1126$-$041:} This system clearly shows a disk-like morphology and rotational pattern traced by the H$\alpha$ emission. A clumpy morphology is seen toward the galaxy outskirts, while in the central zone the H$\alpha$ emission seems to be distributed in a bar-like or outflow structure. The velocity dispersion map is complex, with an enhancement of $\sigma_{V,{\rm H\alpha}}$ toward the north-west and south-east, zones that are ionized by the AGN, although no clear substructures are visible in the H$\alpha$ and [O\,{\sc iii}] intensity maps. Ongoing star formation activity is necessary to explain the BPT maps.

\item[]\textbf{PG\,1211+143}: The MUSE observation shows a compact system in which some evidence of rotation can be appreciated from the LOS velocity map. The data suggest two major clumps toward the north of the galaxy; they are consistent with star formation once the [N\,{\sc ii}] upper limits are considered. The lack of [N\,{\sc ii}] but the detection of H$\alpha$, H$\beta$, and [O\,{\sc iii}] suggests that these two zones are metal-poor. 

\item[]\textbf{PG\,1244+026:} The H$\alpha$ emission traces a compact ionized gas distribution and a star-forming clump present toward the south. Interestingly, while the central emission of the clump is consistently classified as ``composite,'' it is surrounded by pixels with ``Seyfert'' classification according to the [N\,{\sc ii}]/H$\alpha$ map. This suggests that PSF blurring is inducing emission-line flux mixing. A rotation pattern is evident in the LOS velocity map. No clear trend is seen in the velocity dispersion map.

\item[]\textbf{PG\,1426+015:} The host galaxy is an ongoing merger \citep{Kim2017}. Although the H$\alpha$ emission map looks patchy, we can see clearly three different structures: a large tidal feature toward the east, a second tidal feature toward the west, and a central, rotating disk-like component according to the H$\alpha$ LOS velocity map. The BPT maps suggest that star formation activity is probably taking place within the tidal features and in some small clumps located in the central disk-like component. The higher $\sigma_{V,{\rm H\alpha}}$ observed toward the eastern zone of the inner disk-like component may suggest the presence of an outflow component and/or shocks affecting the ionized gas, judging by the [S\,{\sc ii}]/H$\alpha$ diagram.

\item[]\textbf{PG\,2130+099:} The H$\alpha$ emission traces an outer spiral-like clumpy structure that extends out to the outskirts of the galaxy. The clumpy regions are consistently classified as star-forming zones. A central clumpy disk-like region is seen toward the center. The [N\,{\sc ii}]/H$\alpha$ map suggests ongoing star formation, although shocks cannot be completely discarded, according to the [S\,{\sc ii}]/H$\alpha$ map. The LOS velocity maps clearly show a rotating disk-like pattern. However, the velocity dispersion map suggests more complex dynamics. While $\sigma_{V,{\rm H\alpha}}$ increases overall toward the central disk-like region, higher velocity dispersions are seen in the zones between the spiral and the central disk-like region.

\end{itemize}

\vfill\eject
\bibliography{bibliography}
\bibliographystyle{aasjournal}
\end{document}